\documentclass[a4paper, 12pt]{amsart}
\usepackage[foot]{amsaddr}

\usepackage{xr}
\usepackage{amsthm}
\usepackage[mathscr]{eucal}
\usepackage{amsfonts}
\usepackage{amssymb}
\usepackage{graphicx}
\usepackage{caption}
\usepackage{subcaption}
\usepackage{pstricks}
\usepackage{hyperref}
\usepackage{booktabs}
\usepackage{colortbl}
\usepackage{comment}
\usepackage{tabularx}

\usepackage{xcolor}

\newcolumntype{N}{>{\centering\arraybackslash}m{.85in}}

\newrgbcolor{myblue}{0 0 1}
\newrgbcolor{myred}{0.466667 0 0}
\newrgbcolor{mygreen}{0 0.466667 0}
\newrgbcolor{mylightred}{1 0 0}
\newrgbcolor{mylightblue}{0 0.8 1}

\newcommand{\ttlred}[1]{\textcolor{mylightred}{#1}}
\newcommand{\ttlblue}[1]{\textcolor{mylightblue}{#1}}

\usepackage{tabularx,ragged2e,booktabs,caption}
\newcolumntype{C}[1]{>{\Centering}m{#1}}

\newcommand\solidrule[1][0.5cm]{\rule[0.5ex]{#1}{.8pt}}
\newcommand\dashedrule{\mbox{%
  \solidrule[1mm]\hspace{1mm}\solidrule[1mm]\hspace{1mm}\solidrule[1mm]}}

\newcommand\dotsrule{\mbox{%
  \solidrule[0.5mm]\hspace{.45mm}\solidrule[0.5mm]\hspace{.45mm}\solidrule[0.5mm]}}

\usepackage[left=2.5cm, right=2.5cm, top=3.5cm, bottom=3cm]{geometry}
\usepackage[onehalfspacing]{setspace}

\usepackage[authoryear,round]{natbib}

\newcommand{\mb}[1]{\mathbb{#1}}

\newcommand{\bs}[1]{\boldsymbol{#1}}

\newcommand{\msf}[1]{\mathsf{#1}}

\newcommand{\tn}[1]{\textnormal{#1}}

\newcommand{\ind}{{1\hspace{-2.5pt}\tn{l}}}

\DeclareMathOperator*{\argmin}{argmin}

\newcommand{\N}{\mb{N}}
\newcommand{\E}{\mb{E}}

\newcommand{\VAR}{\msf{VAR}}
\newcommand{\SD}{\msf{SD}}
\newcommand{\MSE}{\msf{MSE}}
\newcommand{\COV}{\msf{Cov}}

\newcommand{\BIAS}{\msf{BIAS}}

\newcommand{\kernel}{F}
\newcommand{\sr}{f_{\bs{s}}}

\newcommand{\JULES}{\operatorname{JULES}}
\newcommand{\JSMURF}{\operatorname{JSMURF}}
\newcommand{\TRANSIT}{\operatorname{TRANSIT}}
\newcommand{\MDL}{\operatorname{MDL}}
\newcommand{\valmu}{l}
\newcommand{\median}{\operatorname{median}}

\usepackage{siunitx}
\DeclareSIUnit{\Molar}{\textsc{m}}
\usepackage{algorithm2e}
\usepackage{tikz}
\usetikzlibrary{shapes,arrows,backgrounds,fit,calc,positioning}
\begin{document}

\title[Multiresolution Idealization for Filtered Ion Channel Recordings]{Fully-Automatic Multiresolution Idealization for Filtered Ion Channel Recordings:\\ Flickering Event Detection}

\author{Florian~Pein$^1$, Inder~Tecuapetla-G\'omez$^1,^2$, Ole~Mathis~Sch\"utte$^3$, Claudia~Steinem$^3$ and Axel~Munk$^1,^4,^5$}
\address{$^1$Institute for Mathematical Stochastics, Georg-August-University of G{\"o}ttingen, Goldschmidtstra{\ss}e 7, 37077 G{\"o}ttingen, Germany\newline
$^2$CONACyT-CONABIO-Direcci\'on de Percepci\'on Remota, Lisa Periferico-Insurgentes Sur 4903, Parques del Pedregal, Tlalpan 14010, Mexico D.F.\newline
$^3$Institute of Organic and Biomolecular Chemistry, Georg-August University of Goettingen, Tammannstr.~2, 37077 G\"ottingen, Germany\newline
$^4$Max Planck Institute for Biophysical Chemistry, Am Fa{\ss}berg 11, 37077 G{\"o}ttingen, Germany\newline
$^5$Felix Bernstein Institute for Mathematical Statistics in the Biosciences, Goldschmidtstr. 7, 37077 G\"ottingen, Germany}
\email{\{fpein, itecuap, munk\}@math.uni-goettingen.de}
\email{\{ole.schuette, claudia.steinem\}@chemie.uni-goettingen.de}

\date{\today}

\begin{abstract}
We propose a new model-free segmentation method, JULES, which combines recent statistical multiresolution techniques with local deconvolution for idealization of ion channel recordings. The multiresolution criterion takes into account scales {\black down} to the sampling rate enabling the detection of flickering events, i.e., events on small temporal scales, even below the filter frequency. For such small scales the deconvolution step allows for a precise determination of dwell times and, in particular, of amplitude levels, a task which is not possible with common thresholding methods. This is confirmed theoretically and in a comprehensive simulation study. {\black In addition, $\JULES$ can be applied as a preprocessing method for a refined hidden Markov analysis.} Our new methodolodgy allows us to show that gramicidin A flickering events have the same amplitude as the slow gating events. JULES is available as an \texttt{R} function \textit{jules} in the package \textit{clampSeg}.
\end{abstract}

\maketitle

\section{Introduction}\label{sec:introduction}

\begin{figure}[!htp]
\centering
\includegraphics[width = 0.99\textwidth]{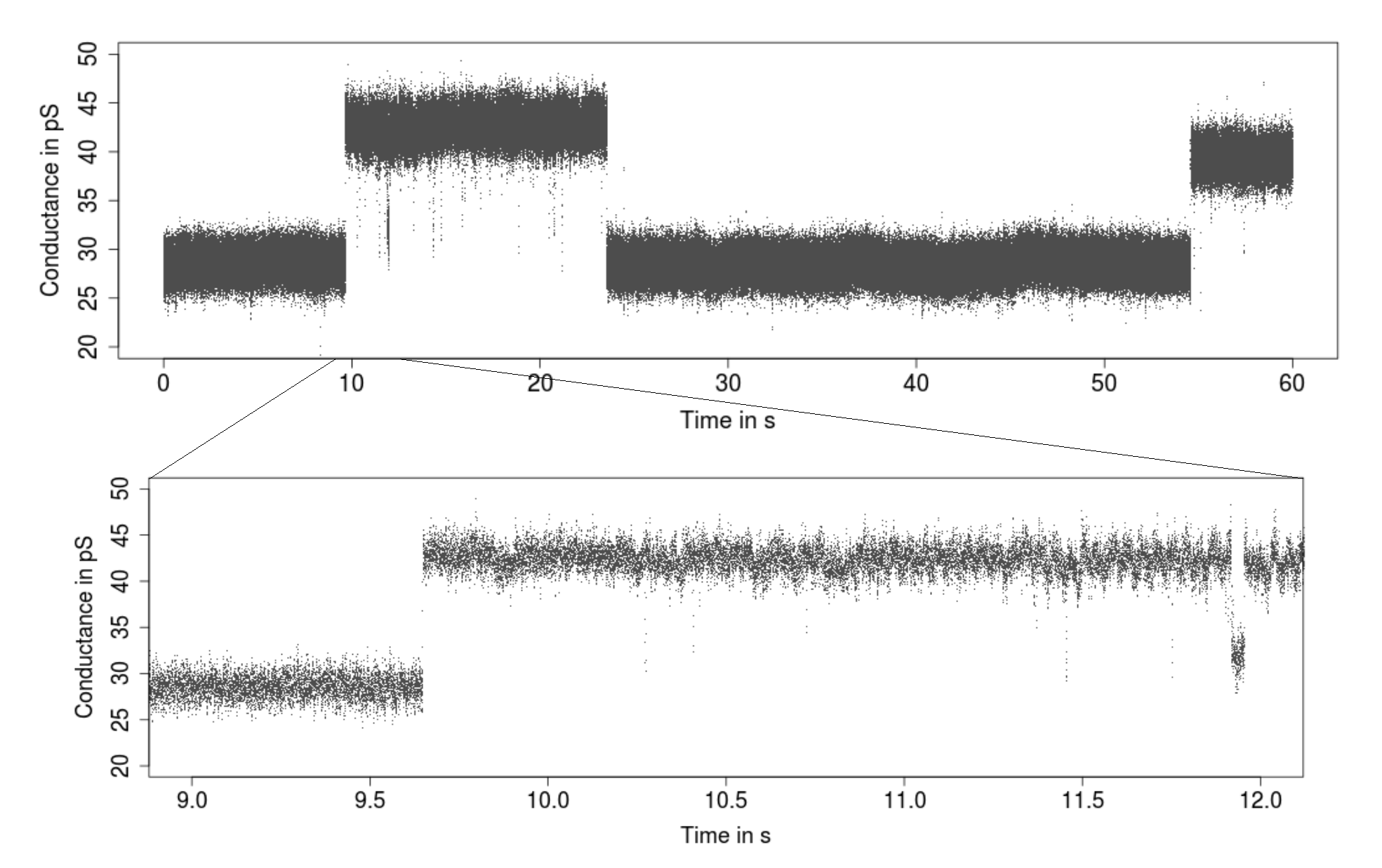}
\caption{Observations (grey points) of a typical conductance time trace obtained for active gramicidin A (gA) inserted into a lipid bilayer.}
\label{fig:GramicidinData}
\end{figure}

\begin{figure}[!htp]
\centering
\includegraphics[width = 0.99\textwidth]{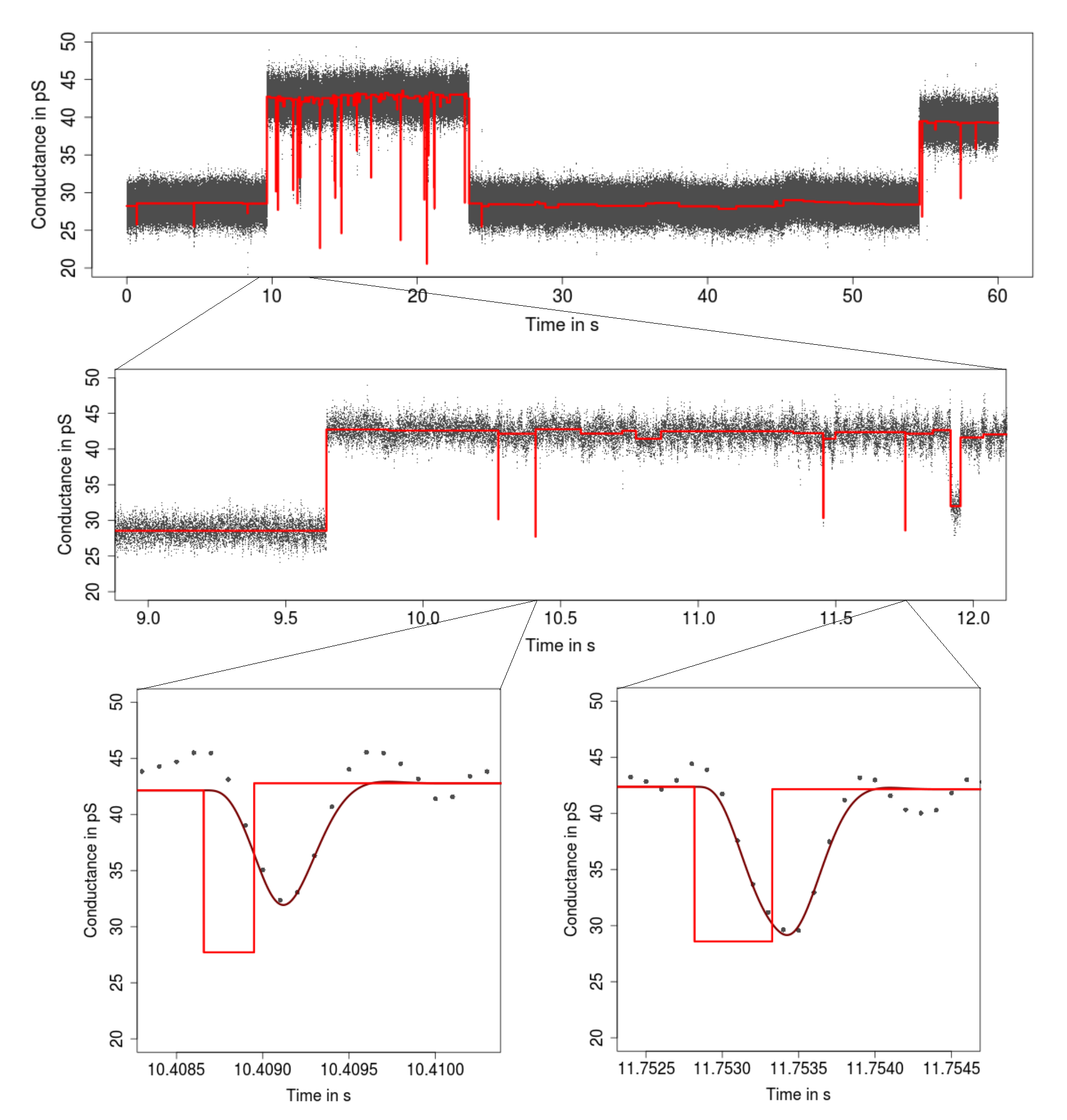}
\caption{Observations (grey points) as in Figure \ref{fig:GramicidinData} together with idealization by $\JULES$ ({\ttlred  \solidrule}) and its convolution with the lowpass filter ({\myred \solidrule}).}
\label{fig:GramicidinJULES}
\end{figure}

\begin{figure}[!htp]
\centering
\includegraphics[width = 0.99\textwidth]{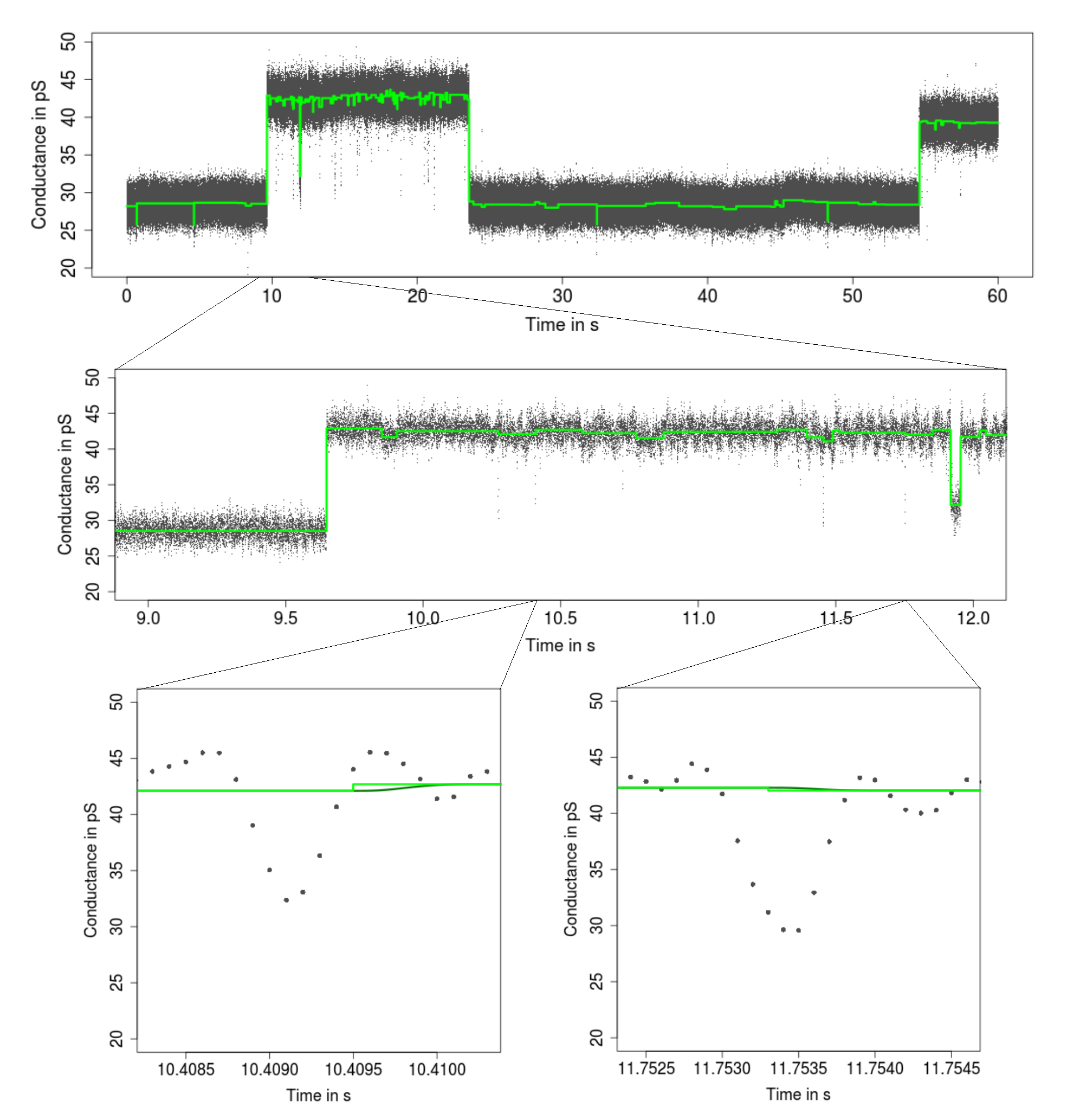}
\caption{Observations (grey points) as in Figure \ref{fig:GramicidinData} together with idealization by $\JSMURF$ ({\green \solidrule}) and its convolution with the lowpass filter ({\mygreen \solidrule}).}
\label{fig:GramicidinJSMURF}
\end{figure}

{\black Ion channels are fundamental to many vital human functions like such as breathing and heartbeating as they allow for a regulative transport of ions through the otherwise ion impermeable lipid membrane along electrochemical gradients. The channels can switch between different conformations (\textit{gating}) resulting in different conductance levels (\textit{states}) controlled by external stimuli such as voltage, ligand binding or mechanical stress \cite{Chung2007biological}. Understanding of ion channel gating has a long history and plays an important role for many purposes, e.g., for developing drugs \cite{kass2005channelopathies, overington2006many}.} A major tool to analyze the gating behavior is the \textit{patch clamp} technique which allows to record the conductance trace (i.e., the recorded current trace divided by the applied voltage) of a \textit{single} ion channel \cite{Neher.Sakmann.76, Sakmann.Neher.95}.\\
{\black To stay in the transmission range of the amplifier the recordings are convolved by a lowpass filter (typically a 4, 6 or 8 pole Bessel filter) and digitized afterwards \cite{Sakmann.Neher.95}. This attenuates high frequent noise components, e.g., caused by shot noise, for an example see Figure \ref{fig:GramicidinData}. Therefore,} lowpass filters are in general integrated in the hardware of the technical measurement device, since they are indispensable for noise reduction and visualization. Important channel characteristics such as amplitudes and dwell times can be obtained provided the conductance changes of the traces are \textit{idealized} (underlying signal is reconstructed) from these recordings (data points) \cite{Colquhoun.87, Sakmann.Neher.95, Hotz.etal.13}.\\
Besides subjective and time-consuming analysis by visual inspection, often with manually chosen event times or in a semi-automatic way, as e.g. offered by pCLAMP 10 software (Molecular Devices), there is a demand for fully automatic methods, which provide idealizations on an objective basis, preferably with statistical guarantees for the recovered events. These automatic methods can be roughly divided into two groups. {\black 
\paragraph*{HMM based methods}
Firstly, methods that assume an underlying statistical model with a few parameters describing the gating dynamics are utilized. Most common are hidden Markov models (HMMs), see \cite{Ball.Rice.92}, where parameters correspond to transition probabilities and levels. These parameters can either be estimated by the Baum-Welch algorithm, see \cite{Venkataramanan.etal.00, Qin.etal.00}, or by MCMC sampling, see \cite{deGunst.etal.01, Siekmann.etal.11}, or by approaches based on the conductance (current) distribution, see \cite{yellen1984ionic, heinemann1991open, schroeder2015resolve} and the references therein. Although, they do not provide an immediate idealization of the underlying signal, the most likely path (idealization) can be obtained from these models, e.g., by the Viterbi algorithm  \cite{viterbi1967error} and related methods. In addition, and relevant to this paper, parameter estimation can often be corrected for missing short events, see \cite{hawkes.etal.90, qin1996estimating, colquhoun1996joint, epstein2016bayesian}.
\paragraph*{Model-free methods} 
Secondly, and more in J. Tukey's spirit of \emph{letting the data speak for itself}, non-parametric, often called \textit{model-free}, methods that do not rely on a specific (parametric) model are used. Examples of such methods are amplitude thresholding \cite{Colquhoun.87, Sakmann.Neher.95}, slope thresholding \cite{Basseville.Benveniste.83, VanDongen.96}, idealization based on the minimal description length ($\MDL$) \cite{gnanasambandam2017unsupervised} and the jump segmentation multiresolution filter $\JSMURF$ \cite{Hotz.etal.13}.\\
The methodology suggested in this paper belongs to this second class of models. We stress, however, that it can also be seen as a companion for HMM based methods, e.g., to determine the number of states and for prefiltering purposes. This will be specified in the following.
\paragraph*{Methods - compared and contrasted}
Usually, gating (\SIrange{1}{100}{\nano\second}) is much faster than the sampling rates (\SIrange{1}{100}{\kilo\hertz}) of the recordings and, hence, channel recordings have the appearance of abrupt random changes between states \cite{hamill.81.etal}. Approaches that assume a HMM, e.g., conductance distribution fitting, allow to resolve the gating dynamics below the measured temporal scales, potentially. Nevertheless, often a rigorous HMM based analysis is hindered by several issues. To specify the Markov model for the gating dynamics it is required to fix the number of states. Although some model selection approaches exist, this is not an easy task, in general. Further, in rare cases the Markovian property itself is questionable, cf.~\cite{Fulinski.98, Mercik.Weron.01, Goychuk.etal.05, Shelley.etal.10} or, more commonly, the implicit homogeneity assumption underlying a hidden Markov model is violated, e.g., due to artifacts providing an inhomogeneous error structure. Hence, in the presence of artifacts like for instance base line fluctuations (cf. Figure \ref{fig:GramicidinData}) the assumption of few conductance levels that occur repeatedly requires elaborate data cleaning before an HMM can be fitted. In contrast, in such situations model-free methods may provide right away a reasonable idealization as they potentially can handle inhomogeneity in a more flexible way. Additionally, the computational complexity of many model-free methods increases almost linearly in the number of data points, an important issue, since often large data sets have to be analyzed. Finally, model-free methods, in addition to model selection approaches, can be used to select or verify a specific Markov model, in particular to determine the number of states and possible transitions, and to explore and potentially remove artifacts. Furthermore, fitting the conductance distribution requires often an initial guess for the parameters which can be provided by model-free methods, as well. Finally, if a Markov model is assumed, missing event corrections can also be combined with model-free methods to improve the parameter estimation based on their idealizations, see Section \ref{sec:hmm}.\\
Consequently, hidden Markov models, fitting the conductance distribution, and model-free methods complement each other well and should not be seen as competing approaches. While model-free methods ease idealization for e.g. nonhomogeneous data, approaches based on a Markov model allows to extrapolate information from larger (observable) to smaller (not observable) time scales.\\
While the literature on hidden Markov models and fitting the conductance distribution is comprehensive, we see a demand for better working model-free approaches and its theoretical justification. Many approaches are based on additional filtering (often by lowpass Gauss filters) and manually chosen thresholds which are difficult to determine in an objective way. It is well known that filtering threatens to miss events in the idealization \cite{Colquhoun.87, Sakmann.Neher.95, Hotz.etal.13} as the filter length predetermines the scales on which events can be detected. Further, such methods often ignore the lowpass filter of the amplifier. But ignoring this filtering threatens to include false positives which falsify the analysis as well.\\
Hence,} providing rigorous statistical guarantees for the idealized traces is lacking to model-free methods, in general, and filtering hinders this additionally. The multiresolution method $\JSMURF$ \cite{Hotz.etal.13} offers a model-free idealization which comes with statistical guarantees for its occurrences of conductance changes at a range of time scales. Roughly speaking, at a given error level $\alpha$ it guarantees detected conductance changes to be significant {\black (controls the false positives)}. However, this method is limited to achieve accuracy at the filter resolution and beyond. As a consequence, it fails to detect \textit{flickering} events ({\black events shorter than the filter length}), see Figure \ref{fig:GramicidinJSMURF} for an illustration and the simulations in Section \ref{sec:simulations}.\\
This is a crucial issue, as flickering typically has its own dynamics and can result from different molecular processes like conformational changes of the ion channel \cite{Grosse.etal.14} or by the passage of larger molecules blocking the ion’s pathway through the channel \cite{Singh.etal.12}. Thus, and to the best of our knowledge, no model-free method is able to detect flickering events reliably, i.e., at a given statistical accuracy, while at the same time detecting events at larger scales. Moreover, even if an event is detected, precise idealization of its amplitude and its dwell times requires new tools, since the convolution of the signal with the filter smoothes the peak, see Figures \ref{fig:GramicidinJULES} and \ref{fig:singlePeak} for illustration and \cite{Colquhoun.87, Sakmann.Neher.95}. {\black An exception is the semi-automatic \textit{SCAN} software \cite{colquhoun1995fitting} which allows \textit{time-course fitting}. This means to idealize events by least squares fitting based on an approximation of the Bessel filter kernel by a Gauss kernel, but interventions by the experimenter are required. Moreover, a conductance value can only be idealized by least squares fitting if the event is long enough, otherwise it is guessed by previously obtained values.}\\
Therefore, to overcome this lack of methodology, in this paper we will present a model-free method for detection and precise idealization of event times and amplitudes of flickering events, while at the same hand all relevant events at larger time scales will be recovered as well.

\paragraph*{JULES}
We first introduce in Section \ref{sec:model} an extremely flexible and simple model for the recordings, which does not rely on a Markov assumption. Then, we propose based on this model a new \textbf{JU}mp \textbf{L}ocal d\textbf{E}convolution \textbf{S}egmentation filter, $\JULES$, that consists of a (model-free) detection step by a \textit{multiresolution constraint} (Section \ref{sec:multiresolution}) and an estimation (idealization) step for the signal by \textit{local deconvolution} (Section \ref{sec:deconvolution}). The level of resolution (encoded in the multiresolution constraint) of $\JULES$ will be of the magnitude of the sampling rate and allows us even detection of events below the filter frequency. To this end, the final idealization of the time-continuous jump locations and in particular of the amplitudes is achieved by local deconvolution for which statistical guarantees can be provided as well.

\paragraph*{Implementation}
These steps are complex and are summarized in Algorithm \ref{alg:jules}. $\JULES$ is implemented as an \texttt{R} \cite{R2016} function \textit{jules} in the package \textit{clampSeg} \cite{clampSeg}. The detection step is computed via a pruned dynamic program implemented in C++ to speed up computations. The deconvolution step is a maximum likelihood estimator computed by an iterative grid search algorithm, see Section \ref{sec:deconvolution}. 


\tikzstyle{line} = [draw, -latex']

\begin{algorithm}
\center
\begin{tikzpicture}[node distance = 1.5cm, auto] \centering
\node [draw,trapezium,trapezium left angle=70,trapezium right angle=-70,minimum height=1cm,fill=green!20] (input) {\shortstack{Data $Y_1,\ldots,Y_n$, error level $\alpha$, regularization parameter $\gamma^2$,\\ filter with kernel $\kernel_m$ truncated at $m$}};

\node[rectangle,fill=blue!20, below of=input] (reconstruction) {Initial reconstruction $\hat{g}$ $\leftarrow$ multiresolution estimator \eqref{eq:multiscaleestimator} by a dynamic program};

\node[rectangle,fill=blue!20, below = 0.5cm of reconstruction] (postfiltering) {Postfiltering to remove incremental steps in $\hat{g}$};

\begin{pgfonlayer}{background} 
\node[minimum width=12.3cm, line width=1pt, dash pattern=on 3pt off 3pt, inner sep=2.5mm, align = center, rounded corners, fit=(reconstruction)(postfiltering), draw] (detection) {};   
\end{pgfonlayer}
\node [right=0.1cm of detection] {\shortstack{\textbf{Detection step}\\ Section \ref{sec:multiresolution}}};

\node[rectangle,fill=blue!20, below of=postfiltering] (deconvolution) {Idealization $\hat{f}$ $\leftarrow$ local deconvolution};

\begin{pgfonlayer}{background} 
\node[minimum width=12.3cm, line width=1pt, dash pattern=on 3pt off 3pt, inner sep=2.5mm, align = center, rounded corners, fit=(deconvolution), draw] (estimation) {};   
\end{pgfonlayer}
\node [right=0.1cm of estimation] {\shortstack{\textbf{Estimation step}\\ Section \ref{sec:deconvolution}}};

\node [draw,trapezium,trapezium left angle=70,trapezium right angle=-70,minimum height=1cm,fill=green!20, below of=deconvolution] (output) {\shortstack{Idealization $\hat{f}$, i.e., all event times and amplitudes}};

\path [line] (input) -- (reconstruction);
\path [line] (reconstruction) -- (postfiltering);
\path [line] (postfiltering) -- (deconvolution);
\path [line] (deconvolution) -- (output);

\end{tikzpicture}
\renewcommand{\thealgocf}{JULES}
\caption{Steps of $\JULES$.}
\label{alg:jules}
\end{algorithm}

\begin{figure}[!htp]
\centering
\includegraphics[width = 0.99\textwidth]{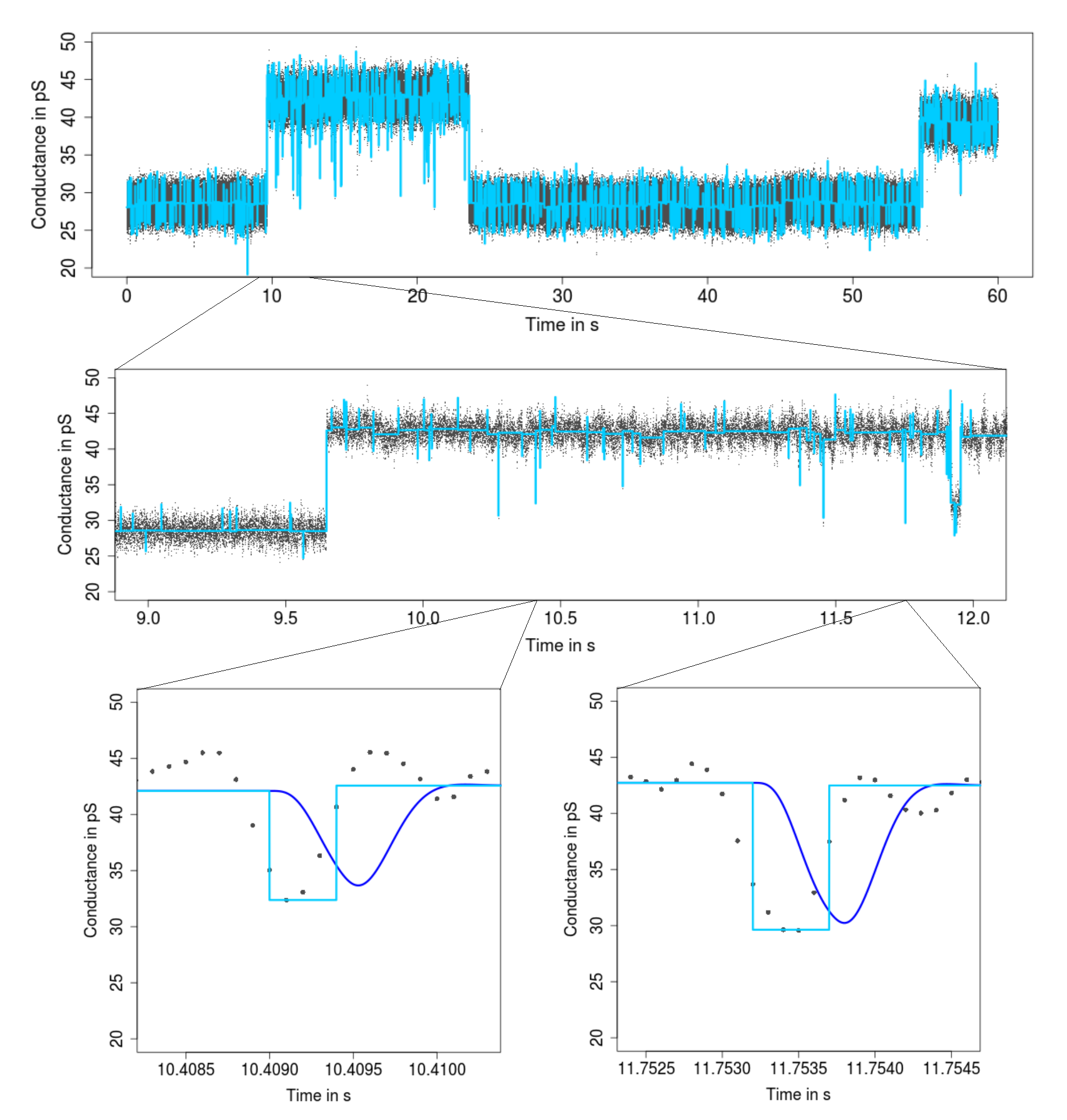}
\caption{Observations (grey points) as in Figure \ref{fig:GramicidinData} together with idealization by $\TRANSIT$ ({\ttlblue \solidrule}) and its convolution with the lowpass filter ({\blue \solidrule}).}
\label{fig:GramicidinTRANSIT}
\end{figure}

\paragraph*{Evaluation}
The performance of $\JULES$ is assessed in several simulation studies (Section \ref{sec:simulations}) and on gramicidin A (gA) traces (Section \ref{sec:analysis}). The simulations indeed confirm that $\JULES$ is able to detect and idealize events that are short in time with very high precision. More precisely, we found that $\JULES$ detects (for the signal-to-noise ratio and filter setup of the later analysed gramicidin A traces) with high probability peaks with length at least one quarter of the filter length, see Figure \ref{fig:GramicidinJULES} for an illustration and Figure \ref{fig:relativeFreqDetection} for a systematic analysis, and separates two consecutive peaks with high probability once the distance between them is at least three times the filter length, see Figure \ref{fig:peakSeparation}. Although not designed under a Hidden Markov assumption, Hidden Markov model simulations with two flickering states show that $\JULES$ is able to separate them if the amplitude of the two states differs at least by $\SI{3}{\pico\siemens}$ (which is roughly two times the noise level and one sixth of the amplitude). In comparison, $\JSMURF$ misses all peaks below the filter length, see Figure \ref{fig:GramicidinJSMURF} for illustration, whereas $\TRANSIT$ \cite{VanDongen.96} includes many additional wrong detections, illustrated in Figure \ref{fig:GramicidinTRANSIT}. Moreover, its idealization shifts peaks to the right and estimates too small amplitudes for shorter lengths when the signal is smoothed by the filter, see Figure \ref{fig:singlePeak_b} and Table \ref{tab_singlePeak_estimationLevel}.\\
{\black JULES can be nicely combined with HMMs: If in addition the gating dynamics are assumed to follow a Markov model, even shorter dwell times can be analyzed by using missed event corrections. More precisely, for the simulated signal to noise ratio and the used filter setup, dwell times slightly below the sampling rate ($10$ kHz) are analyzable by $\JULES$, since then still enough events are long enough to be detected. As discussed before, methods that directly use the Markov model, in particular conductance distribution fitting, might be able to resolve faster gating, however at the price of for instance a much smaller robustness against model violations.}

\paragraph*{Data analysis}
Gramicidn A is a small peptide produced by the soil bacterium \textit{Bacillus brevis} showing antimicrobial activity \cite{Connell.etal.90, Andersen2007gramicidin}.  It forms an ion channel by interleaflet dimerization making its putative gating mechanism sensible to properties of the surrounding lipid membrane \cite{Connell.etal.90, andersen2005gramicidin, Andersen2007gramicidin}. Apart from the classical slow gating by dimerization also short closing events of the channel have been observed which could be linked to the physical and chemical properties of the lipid membrane; especially its hydrophobic thickness \cite{Armstrong2002origin, Ring1986brief}.\\
Idealization of the gramicidin A traces supports the finding of the simulations with high evidence, see Figures \ref{fig:GramicidinJULES}-\ref{fig:GramicidinTRANSIT}. Zooming into single peaks demonstrates that $\JULES$ idealizes the observation extremely well (see Figure \ref{fig:GramicidinJULES} lower panels), which also supports our underlying model. As all simulations (see above) haven been performed for models which are resembling our gramicidin A data sets very well, this establishes $\JULES$ as a reliable tool to identify and idealize flickering events in filtered ion channel recordings. We found that the flickering events are full-sized, i.e., have the same amplitude than the slow gating events, see Figure \ref{fig:histlevel}. In the near future we will continue with an analysis of further transmembrane proteins. Currently we examine for instance whether certain mutations influence the transport of antibiotics in a bacterial porin and analyze the gating dynamics of the translocase channel Tim23.

\section{Modeling}\label{sec:model}
As argued above (see also \cite{Hotz.etal.13}) the conductivity will be modeled as a piecewise constant signal
\begin{equation}\label{eq:signal}
    f(t) = \sum_{j=0}^{K}\,\valmu_j\,\ind_{[\tau_j, \tau_{j+1})}(t),
\end{equation}
where $t$ denotes physical time. The (unknown) levels are denoted as $\valmu_0, \ldots, \valmu_K$, its (unknown) number of conductivity changes as $K$ and the (unknown) changes between states as ${\black -\infty} =: \tau_0 < \tau_1 < \cdots < \tau_K < \tau_{K+1}:=\tau_{\operatorname{end}}$. {\black  Note, this implies that we only consider conductivity changes after recording started at time $0$ and assume for simplicity that the signal before recordings started is constant and equal to the first function value. When truncating the filter we will see that only a very short time period before recordings started will be relevant, details are given at the end of this section. For an analysis of how long the channel dwells in the same conductance level we exclude always the first and last segment. We stress that the class of signals in \eqref{eq:signal} is very flexible as potentially an arbitrary number of states and changes at arbitrary levels can be imposed, see again Figure \ref{fig:GramicidinJULES} for an example.} We assume further that the recorded data points $Y_1,\ldots,Y_n$ (the measured conductivity at time points $t_i=i/\sr,\ i = 1,\ldots,n$, equidistantly sampled at rate $\sr$) result from convolving the signal $f$ perturbed by gaussian white noise $\eta$ with an analogue lowpass filter, with kernel $\kernel$, and digitization at sampling rate $\sr = n / \tau_{\operatorname{end}}$, i.e.,
\begin{equation}\label{eq:model}
Y_i = \big(\kernel \ast (f + \sigma_0 \eta)\big)\left(i/\sr\right) = (\kernel \ast f)(i/\sr) + \epsilon_i,\quad i=1,\ldots,n,
\end{equation}
with noise level $\sigma_0>0$ and $\ast$ the convolution operator. Here $n$ denotes the total number of data (typically several hundred thousands up to few millions). Hence, the resulting errors $\epsilon_1,\ldots,\epsilon_n$ are gaussian and centered, $\E[\epsilon_i]=0$, and have covariance $\COV[\epsilon_i,\epsilon_{i+j}]=\sigma_0^2(\kernel \ast \kernel)(j/\sr)$. For example the gA traces in our subsequent analysis (see Figure \ref{fig:GramicidinJULES} and Section \ref{sec:analysis}) are filtered by a 4-pole lowpass Bessel filter with $\SI{1}{\kilo\hertz}$ cut-off frequency and sampled at $\SI{10}{\kilo\hertz}$. The major aim will be now to idealize (reconstruct) the unknown signal $f$ taking into account the convolution and the specific structure of $f$ in \eqref{eq:signal}. This will be done fully automatically and with statistically error control. By fully automatic we mean that no user action is required during the idealization process, only the error level $\alpha$ and two filter specific parameters have to be selected in advance, see Section \ref{sec:parameterchoices}. This is no restriction when the filter is known (as it is the case in most applications) and when it is unknown it can be pre-estimated from the data \cite{Munk.Tecuapetla.15}.\\
Typically used lowpass filters have infinite impulse responses but a quickly decaying kernel $\kernel$, for instance the kernels of Bessel filters decay exponentially fast \cite{Proakis.Manolakis.96}. Thus, we truncate the kernel $\kernel$ at some point, say $m{\black / \sr}$, which provides further simplification of our model. The trucated kernel is denoted by $\kernel_m$ and rescaled such that $\int_{-\infty}^{\infty}{\kernel_m}=1$, see \cite{Hotz.etal.13}. This is implemented in the \texttt{R} function \textit{lowpassFilter} \cite{clampSeg}. We also truncate the autocorrelation function of the noise, resulting in an $m$-dependent process. All in all, model \eqref{eq:model} reduces to
\begin{equation}\label{eq:modeltrun}
Y_i = (\kernel_m \ast f)(i/\sr) + \epsilon_i,\quad i=1,\ldots,n,
\end{equation}
where $\epsilon_1,\ldots,\epsilon_n$ have covariance
\begin{equation}\label{eq:covtrun}
\COV\big[\epsilon_i,\epsilon_{i+j}\big]=\left\{\begin{array}{cl}
\sigma_0^2(\kernel \ast \kernel)(j/\sr), & \text{for } \vert j \vert=0,\ldots,m,\\
0, & \text{for } \vert j\vert > m.
 \end{array}\right.
\end{equation}
Note that we truncate the autocorrelation function itself as this leads to more accurate values for lags close to $m$ instead of convolving the truncated kernel. As a working rule, we choose $m$ such that the autocorrelation function ${\black(\kernel \ast \kernel)(\cdot)/(\kernel \ast \kernel)(0)}$ of the analogue lowpass filter is below $10^{-3}$. For gA traces, this choice leads to $m = 11$ and is confirmed visually by comparing the observations with the convolution of the kernel with the idealized signal, see Figures \ref{fig:GramicidinJULES} and \ref{fig:singlePeak}.

\section{Methodology: JULES}\label{sec:methodology}
\subsection{Detection step: multiresolution reconstruction}\label{sec:multiresolution}
In this section we present our novel multiresolution methodology for model \eqref{eq:modeltrun} that detects all significant conductance changes in the recordings and allows us, at the same time, to gain precise statistical error control of the reconstructed quantities. One major challenge is to take temporal scales below the filter frequency into account. We begin by disregarding for the moment the underlying convolution of the signal with the lowpass filter and developing a reconstruction method for the model
\begin{equation}\label{eq:modelmultiresolution}
Y_i = g(i/\sr) + \epsilon_i,\quad i=1,\ldots,n.
\end{equation}
Here $g$ is also a function of type \eqref{eq:signal}, which may have different changes and levels. In other words, with respect to model \eqref{eq:modeltrun}, in a first step we aim for a piecewise constant approximation of $\kernel_m \ast f$ which will then be further refined at the end of this section via postfiltering to a reconstruction of $f$. {\black These two steps fix the number of conductance changes and their rough location.} Afterwards, the final idealization $\JULES$ of $f$ will be obtained by local deconvolution in Section \ref{sec:deconvolution}. A summary of these steps is given in Algorithm \ref{alg:jules}.\\
Observe that with \eqref{eq:modelmultiresolution} we intend to model an unknown piecewise constant function which is perturbed by an $m$-dependent (colored) noise. Its correlation structure, however, is given by the filter and therefore explicitly known, see \eqref{eq:covtrun}. Alternatively, it could be preestimated as in \cite{Munk.Tecuapetla.15}. We preestimate the noise level $\sigma_0$ by
\begin{equation}\label{eq:estsigma0}
\hat{\sigma}_0=\frac{\operatorname{IQR}\left(Y_{m+1}-Y_1,\ldots, Y_n-Y_{n-m}\right)}{2\Phi^{-1}(0.75)\sqrt{2(\kernel \ast\kernel)(0)}},
\end{equation}
used in \cite{Hotz.etal.13} and implemented in the \texttt{R} function \textit{sdrobnorm} \cite{Hotz.Sieling.15}. At this point (a refinement will be provided in the local deconvolution step) we restrict all change-point locations to the grid on which the observations are given, in other words, we assume that $\sr\tau_i$ are integers.

\paragraph*{Multiresolution statistics}
Suppose that we are given a piecewise constant candidate $\tilde{g}$ from model space \eqref{eq:signal} and we have to decide whether or not $\tilde{g}$ is a good fit (reconstruction) for the unknown signal $g$. To this end, we \emph{test simultaneously} on all scales (resolution levels) whether $\tilde{g}$ fits the data appropriately. More precisely, $\tilde{g}$ is a suitable candidate if 
\begin{equation}\label{eq:multistats}
  M(\tilde{g}) := 
 \max_{\substack{1\leq i \leq j \leq n\\
 \tilde{g}([i/\sr,j/\sr]) \text{ const.}}}
 \,\left\{\vert S_{i,j} \vert + p_{i,j}\right\}\\
  := 
 \max_{\substack{1\leq i \leq j \leq n\\
 \tilde{g}([i/\sr,j/\sr]) \text{ const.}}}\,\left\{\left\vert \frac{1}{\sigma_{i,j}}\,\sum_{k=i}^{j}\,\big( Y_k - \tilde{g}(k/\sr) \big) \right\vert + p_{i,j}\right\} \leq q,
\end{equation}
where $p_{i,j}=\sqrt{2\log(e (j - i + 1)/n)}$ is a penalty term which is used to set different scales on equal footing, $\sigma_{i,j} := \sqrt{\VAR( \epsilon_i + \cdots + \epsilon_{j} )} = \sigma_0\,\sqrt{ (j-i+1)(\kernel \ast \kernel)(0) + 2\,\sum_{k=1}^{m}{(j-i+1-k)_+(\kernel \ast \kernel)(k/\sr)}}$ and $q=q(\alpha)$ is a critical value that can be obtained in a universal manner by Monte Carlo simulations such that \eqref{eq:multistats} is a level $\alpha$-test. Here $\alpha$ is fixed in advance (by the experimenter) to control the number of fluctuations of the solution $\tilde{g}$. For details how to generate observations from model \eqref{eq:modeltrun} to simulate $q$, see Section \ref{sec:simulations}. However, to keep it computationally feasible we only {\black evaluate the maximum in \eqref{eq:multistats} over the system of all intervals that contain a dyadic number of observations}, i.e., the maximum in \eqref{eq:multistats} is only taken over all $1\leq i \leq j\leq n$ such that $j - i + 1 = 2^l$ for some $l\in\N_0$. {\black This reduces the complexity of the system from $n(n+1)/2\in\mathcal{O}(n^2)$ to $\lfloor \log_2(n)\rfloor (2n-\lfloor \log_2(n)\rfloor+1)/2\in \mathcal{O}(n\log(n))$ intervals.}\\
In order to provide a reconstruction for $g$ we proceed as follows. Let $G_k$ denote the class of all candidates in \eqref{eq:signal} with $K=k$ changes, $k\geq 0$, and let $\hat{K}$ denote the minimal $k$ for which there exits a $\tilde{g}\in G_k$ such that $M(\tilde{g}) \leq q$, cf.~\eqref{eq:multistats}. Then, the reconstruction for $g$ is given by
\begin{equation}\label{eq:multiscaleestimator}
  \hat{g}(t) := \sum_{j=0}^{\hat{K}}\,\hat{\valmu}_j\,\ind_{[\hat{\tau}_j, \hat{\tau}_{j+1})}(t) 
  := \argmin_{\tilde{g}\in G_{\hat{K}}, M(\tilde{g})\leq q}\,\sum_{i=1}^n\big( Y_k - \tilde{g}(k/\sr) \big)^2.
\end{equation}
Such an estimator has many favorable properties, for instance it detects changes at the optimal detection rate and allows for confidence statements. For more details to the theory behind multiresolution tests as in \eqref{eq:multistats} {\black and the choice of the penalty} see e.g. \cite{Siegmund.Yakir.00, Dumbgen.Spokoiny.01, Dumbgen.Walther.08} and for estimates of type \eqref{eq:multiscaleestimator} see \cite{boysen2009consistencies, Frick.Munk.Sieling.14} and the references therein.

\paragraph*{Dynamic Programming}
The estimator in \eqref{eq:multiscaleestimator} can be computed by {\black dynamic} programming as described in section 3 of \cite{Frick.Munk.Sieling.14}. For related ideas, see also \cite{killicketal12, li2016fdr, maidstone2017optimal} and the references given there. However, its {\black worst case} computation time is quadratic in the number of observations. In order to speed up computations we included several pruning steps in the backward as well as in the forward step of the dynamic program, for more details see section A.1 in the supplement of \cite{Pein.Sieling.Munk.16}. In the situation of routinely changes (which is typical for patch clamp recordings) the runtime of such a program is almost linear in the number of observations such that typical traces (several hundred thousands up to several million data points) can be processed in less than a minute on a standard laptop. An implementation of $\JULES$ (including the local deconvolution step, see next section) can be found in the CRAN-\texttt{R} package \textit{clampSeg} \cite{clampSeg}. 

\paragraph*{Postfiltering}
Our preliminary reconstruction disregarded that also the signal is convolved with the filter and hence we obtained a piecewise constant approximation on the convolution $\kernel_m \ast f$ instead of a reconstruction of the signal $f$ itself. However, they only differ at the beginning of each segment, i.e., when $f$ changes abruptly at $\tau$ from $\valmu_0$ to $\valmu_1$, then $\kernel_m \ast f$ changes on $(\tau, \tau + m / \sr)$ from $\valmu_0$ to $\valmu_1$ and is constant afterwards (as long as no other change occur). As a consequence the reconstruction $\hat{g}$ in \eqref{eq:multiscaleestimator} potentially changes in \emph{incremental} steps, see \cite{li2017multiscale} for a mathematically rigorous statement. To correct for this we merge a segment with all subsequent segments for which the distance between the two starting points is less than $m/\sr$ and if all changes are in the same direction. More precisely, we set the starting point of the merged segments to the starting point of the first segment and its level to the level of the last segment.
 
\subsection{Estimation step: local deconvolution}\label{sec:deconvolution}

In this parameter estimation step we obtain the final idealization by determining the precise changes $\tau_1,\ldots,\tau_{\hat{K}}$ and levels $\valmu_0,\ldots,\valmu_{\hat{K}}$ by local deconvolution. This step is crucial for idealizing (reconstructing) flickering events with high accuracy. To this end, we split the constant segments in short and long ones. Basically, long segments are those in which we can determine its level without taking the lowpass filter explicitly into account, details are given below. Pre-estimating the level of long segments enables to deconvolve the signal \textit{locally}, i.e., to compute the (regularized) maximum likelihood estimator of the remaining parameters (change-point locations and levels) by an iterative grid search, see below for algorithmic details.

\paragraph*{Long segments}
We will define a segment $[\hat{\tau}_i,\hat{\tau}_{i+1})$ as long if we can determine its level $\hat{\valmu}_i$ without taking the lowpass filter explicitly into account. If the signal $f$ changes at $\tau_i$ its convolution $\kernel_m\ast f$ will change continuously on $(\tau_i,\tau_i + m / \sr)$ and will be constant on $[\tau_i + m / \sr, \tau_{i+1}]$. Hence, the detection step will (if the change is detected) find a change in the interval $[\tau_i, \tau_i + m / \sr]$. The other way around, a detected change $\hat{\tau}_i$ implies a change in $\hat{f}$ in the interval $[\hat{\tau}_i - m / \sr,\hat{\tau}_i]$ and, hence, a constant convolution $\kernel_m\ast \hat{f}$ on $[\hat{\tau}_i + m / \sr,\hat{\tau}_{i+1} - m / \sr]$. If the observations in this interval are enough to determine the level, say ten or more, by the median, i.e., $\hat{\valmu}_i=\median(Y_{\sr \hat{\tau}_i + m},\ldots, Y_{\sr \hat{\tau}_{i+1} - m})$, we define the segment as long.

\paragraph*{Local deconvolution}
Knowing these levels enables us to perform the deconvolution locally, i.e., only few observations have to be taken into account and only few parameters have to be estimated. More precisely, let $[\hat{\tau}_{i - 1},\hat{\tau}_{i})$ and $[\hat{\tau}_j,\hat{\tau}_{j+1})$, $i\leq j$, be two consecutive long segments (no long segment inside, but potentially some short segments). Then, we aim to determine the precise locations of the changes $\hat{\tau}_{i},\ldots,\hat{\tau}_{j}$ and the precise levels $\hat{\valmu}_{i},\ldots,\hat{\valmu}_{j-1}$. Note that the levels $\hat{\valmu}_{i-1}$ and $\hat{\valmu}_{j}$ are already determined. And recall that a detected change-point $\hat{\tau}$ implies a change in $[\hat{\tau} - m / \sr,\hat{\tau}]$ and that the level $\hat{\valmu}_i$ affects the convolution on $(\hat{\tau}_i,\hat{\tau}_{i+1} + m / \sr)$. Therefore, we only have to take into account the observations $Y_{\sr \hat{\tau}_i - m + 1},\ldots,Y_{\sr \hat{\tau}_j + m - 1}$ and maximize its likelihood in the unknown parameters which is equivalent to minimize
\begin{equation}\label{eq:likelihood}
(Y_{i,j}-\hat{\mu}_{i,j})^T\Sigma_{i,j}^{-1}(Y_{i,j}-\hat{\mu}_{i,j}),
\end{equation}
with $Y_{i,j}:=(Y_{\sr \hat{\tau}_i - m + 1},\ldots,Y_{\sr \hat{\tau}_j + m - 1})^T$ the vector of the observations, 
\begin{equation*}
\hat{\mu}_{i,j}=\big((K\ast\hat{f})((\sr \hat{\tau}_i - m + 1)/n),\ldots,(K\ast\hat{f})((\sr \hat{\tau}_j + m - 1)/n)\big)^T
\end{equation*}
the vector of the expected conductance levels $K\ast\hat{f}$ for a candidate signal $\hat{f}$ and $\Sigma_{i,j}$ the (known) correlation matrix of the observation vector, see the explanations after \eqref{eq:modeltrun}. {\black For the signal $\hat{f}$ the number of change-points and the function values $\hat{\valmu}_{i-1}$ and $\hat{\valmu}_{j}$ are fixed by the prior reconstruction and the change-point locations are restricted to the intervals $[\hat{\tau}_k - m / \sr,\hat{\tau}_k]$, $k=i,\ldots,j$.} The minimization is performed by an iterative grid search, see the next but one paragraph. 

\paragraph*{Regularization}
For the Bessel filter used in the later analyzed gA traces we have to regularize the correlation matrix $\Sigma_{i,j}$ in \eqref{eq:likelihood}, since the matrix is ill-conditioned, i.e., {\black has a condition number around $10^{-4}$}. This characteristic is typical for most analogue lowpass filters. Hence, small errors in the model (for instance in ion channel recordings often base line fluctuations occur) will be amplified {\black by inversion of $\Sigma_{i,j}$} and may result in a large missestimation. A standard approach to cope with this is \textit{Tikhonov / $L_2$} regularization, i.e., we replace $\Sigma_{i,j}$ by $\Sigma_{i,j} + \gamma^2 I$, with $I$ the identity matrix and $\gamma^2\geq 0$ a regularization parameter, see e.g. \cite[chapter 5]{engl2000regularization}. Such a regularization allows quick computation of \eqref{eq:likelihood} by the Cholesky decomposition (which can be stored) and by solving a triangular system of equations. Note that the correlation structure is determined by the filter and hence the parameter $\gamma^2$ can be chosen universally for all recordings on the same system. For our data we choose the regularization parameter such that the convolution fits well to the recordings, which is nicely confirmed by visual inspection as illustrated in Figure \ref{fig:GramicidinJULES}. This precalibration step has to be done only on a small data excerpt in advance. We found that the results are robust in $\gamma^2$, and simply setting $\gamma^2 = 1$ led to satisfactory results in all cases. For automatic choices, e.g. by cross-validation, and for other regularization, e.g. truncated SVD, see \cite{vogel.02} and references therein.

\paragraph*{Iterative grid search}
In this paragraph we describe how we compute the (local) maximum likelihood estimator by an iterative grid search. In general, computing the estimator, i.e., performing the deconvolution, is difficult due to the non-convexity of the optimization problem to minimize \eqref{eq:likelihood}. Grid search means that we fix for each change $\hat{\tau}_i,\ldots,\hat{\tau}_j$ a set of possible locations, the grid, and compute \eqref{eq:likelihood} for all combinations and take the solution with the minimal value. Note that for given changes close formulas for the optimal levels $\hat{\valmu}_{i},\ldots,\hat{\valmu}_{j - 1}$ exist which allows fast computation of \eqref{eq:likelihood}. In theory such an optimization could be done for arbitrarily fine grids. However, to keep it computationally feasible, we start with the observations grid, i.e., $\{\hat{\tau}_k - m / \sr, \ldots, \hat{\tau}_k\}$ for the change $\hat{\tau}_k, k = i, \ldots,j - 1$. Recall that a detected change $\hat{\tau}_i$ implies a change in $f$ in the interval $[\hat{\tau}_i - m / \sr,\hat{\tau}_i]$. Afterwards, a refinement can be done by repeating the grid search iteratively with finer and finer grids around the changes found in the previous step. We found that this refinement for determining the changes with arbitrarily precision works very well in practice. We found it sufficient to iterate such a refinement twice, each with a ten times finer grid between the neighboring grid points of the candidate for the location of the change.\\
This approach is computationally feasible for a jump between two long segments, i.e., only the jump location is unknown, and for a peak, i.e., a short segment between two long segments, here the unknowns are the level of the short segment and the two jump locations of the short segment between two long segments. See Figures \ref{fig:GramicidinJULES} and \ref{fig:singlePeak} for illustration. In rare situations more than one short segment between two long segments occur. The same approach could then be done for $k$ consecutive short segments by looking for $k+1$ changes between two long segments and determining all $k$ local levels. However, such an optimization is time consuming and not necessary for our data. Hence, we abandon and simply drop these segments from the analysis. For the idealization we keep the reconstruction from the detection step. This leads to potentially miss of events if they are too close together. Figure \ref{fig:peakSeparation} shows that for the analyzed gramicidin A traces this only occurs if the distance between two events is less than \SI{3.2}{\milli\second} (roughly $3$ times the filter length). This is much smaller than estimated average distance between two events of $\SI{0.3}{\second}$ and hence the whole effect seems indeed negligible.

\section{Simulations}\label{sec:simulations}
The purpose of this section is threefold. We will assess first the ability of $\JULES$ in detecting and idealizing an isolated peak. Then, we will identify the minimal distance at which $\JULES$ is able to separate two consecutive peaks. Finally, although $\JULES$ does not rely on a Hidden Markov model assumption we will examine the ability of $\JULES$ to recover the parameter of a Hidden Markov model with one open state and two closed flickering states, in particular how well the two flickering states can be separated. We will generate all signals and observations accordingly to \eqref{eq:signal} and \eqref{eq:modeltrun} such that they are in line with the data we analyze in Section~\ref{sec:analysis}. We also choose amplitudes, dwell time and noise level of the generated observations such that they are similar to those of the datasets of our applications and we also simulate a 4-pole Bessel filter with \SI{1}{\kilo\hertz} cut-off frequency and sample the observation at \SI{10}{\kilo\hertz}. For data generation we use the Durbin-Levinson algorithm \cite[proposition 5.2.1]{Brockwell.Davis.06} to compute the coefficients of the moving average process corresponding to the covariance of the errors in \eqref{eq:modeltrun} and compute the convolution of the simulated signal with the truncated kernel $\kernel_m$ exactly. {\black We also examine briefly robustness issues against $f^2$ and $1/f$ noise.}\\
For comparison, we also apply the slope thresholding method $\TRANSIT$ \cite{VanDongen.96} and the jump segmentation multi\-re\-so\-lu\-tion filter $\JSMURF$ \cite{Hotz.etal.13}. To this end, we used the \texttt{R} functions \texttt{transit} (with default parameters) and \texttt{jsmurf} (also with the default interval system of dyadic lengths and $\alpha = 0.05$) in the package \textit{stepR} \cite{Hotz.Sieling.15}. {\black We also briefly investigate the performance of $\MDL$ \cite{gnanasambandam2017unsupervised} for which we used the Matlab code provided by the authors.}

\subsection{Parameter choices}\label{sec:parameterchoices}
The following specifications will be used in this and in the upcoming section where we analyze the gA traces. For the detection step (multiresolution reconstruction) of $\JULES$ we aim to control the probability to overestimate the number of change-points at significance level $\alpha$ and hence use a conservative choice of $\alpha = 0.05$, resulting in a quantile of $q=1.4539$. In the estimation step (local deconvolution) of $\JULES$ we use a \emph{regularized} correlation matrix with regularization parameter $\gamma^2=1$, see Section \ref{sec:deconvolution}. And, as mentioned before, we truncate the kernel and autocorrelation function of the filter at $m=11$ which corresponds to the fact that the autocorrelation function is below $10^{-1}$ afterwards. These choices are used as default by the \texttt{R} function \textit{jules} in the package \textit{clampSeg} \cite{clampSeg}.

\subsection{Isolated peak}~\label{sec:singlePeak}
\begin{figure}[htb]

  \begin{subfigure}[b]{0.32\linewidth}
    \centering
    \includegraphics[width=\linewidth]{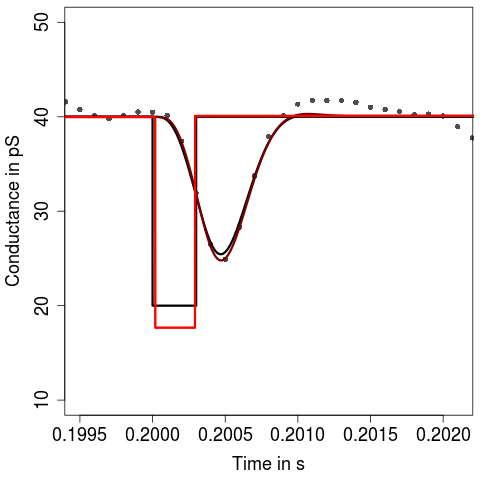} 
    \caption{$\JULES$} 
    \label{fig:singlePeak_a} 
    \vspace{4ex}
  \end{subfigure}
  \begin{subfigure}[b]{0.32\linewidth}
    \centering
    \includegraphics[width=\linewidth]{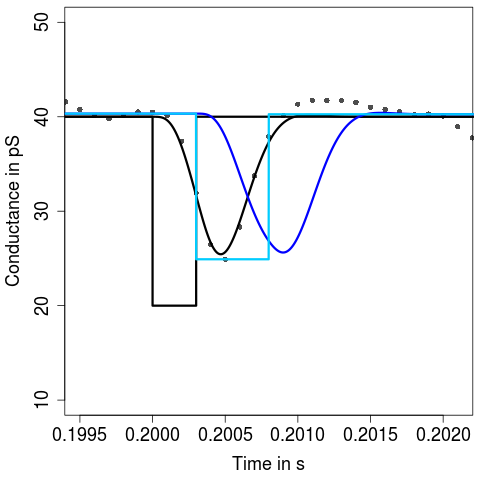} 
    \caption{$\TRANSIT$} 
    \label{fig:singlePeak_b} 
    \vspace{4ex}
  \end{subfigure}
    \begin{subfigure}[b]{0.32\linewidth}
    \centering
    \includegraphics[width=\linewidth]{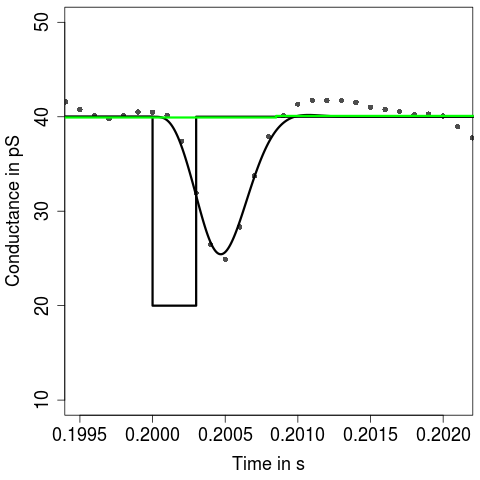} 
    \caption{$\JSMURF$} 
    \label{fig:singlePeak_c} 
    \vspace{4ex}
  \end{subfigure}  
  \vspace{-.5cm}
\caption{\footnotesize Simulated observations (grey points), true block signal $f$ ({\black \solidrule}) and its convolution ({\black \solidrule}), $\JULES$ ({\ttlred  \solidrule}), $\TRANSIT$ ({\ttlblue \solidrule}) and $\JSMURF$ ({\green \solidrule}) idealizations. Convolution of $\JULES$ ({\myred \solidrule}) and $\TRANSIT$ ({\blue \solidrule}) idealizations with the lowpass 4-pole Bessel filter. $\JULES$ provides very accurate idealization, whereas $\TRANSIT$ shifts to the right and estimates a too small amplitude for smaller lengths and $\JSMURF$ misses such peaks.}
  \label{fig:singlePeak} 
\end{figure}

\begin{figure}[htb]
  \centering 
    \includegraphics[width=.99\textwidth]{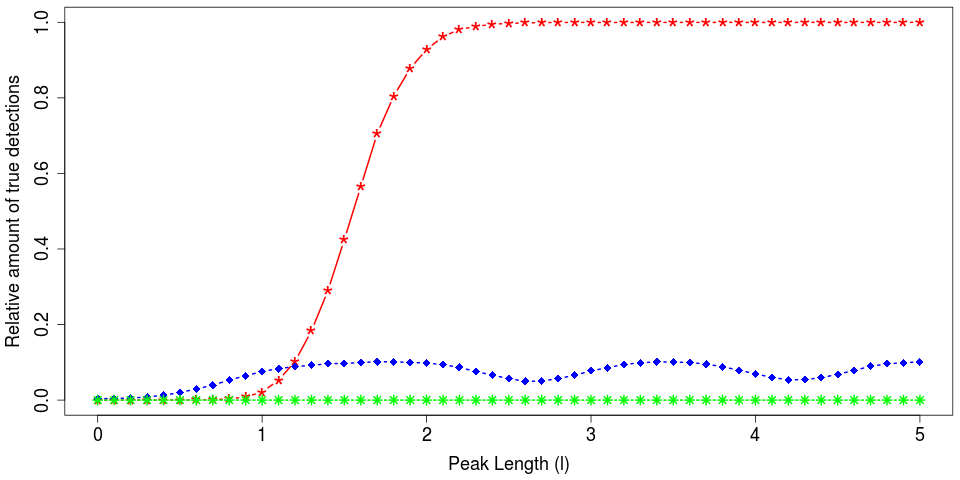} 
    \caption{\footnotesize{Correct identification rates of $\JULES$ ({\mylightred $\star$}), $\TRANSIT$ ({\blue $\blacklozenge$}) and $\JSMURF$ ({\green \textasteriskcentered}) for an isolated peak with length $\ell$. Curves are based on 10\,000 repetitions, each.}}
    \label{fig:relativeFreqDetection} 
\end{figure}

In this simulation with $4\,000$ observations we examine the detection and idealization of a single isolated peak. We consider a situation that is comparable to the analyzed data in Section \ref{sec:analysis}. More precisely, we simulate from model \eqref{eq:modeltrun} with a signal with levels $l_0=l_2=40$ and $l_1=20$ and changes at $\tau_1=2000 / \sr$ and $\tau_2 = ( 2000 + \ell) / \sr$, c.f.~\eqref{eq:signal} and Figure~\ref{fig:singlePeak}. The standard deviation of the errors $\epsilon_i$ in \eqref{eq:modeltrun} used in each simulation is $\sigma_0 = 1.4$. We are interested in the performance of the methods in detecting the peak and estimating the change-point locations $\tau_1$ and $\tau_2$ and the level $l_1$ as a function of $\ell$, the length (relative to the sampling rate $\sr$) of the peak.\\
Tables~\ref{tab_singlePeak_detection}-\ref{tab_singlePeak_estimationLevel} summarize our results based on $10\,000$ repetitions for $\ell=2,3,5$. For $\ell=5$, Figure~\ref{fig:singlePeak} shows an example of the simulated data as well as the idealizations by $\JULES$, $\TRANSIT$ and $\JSMURF$ and their convolutions with the Bessel filter in a neighborhood of the peak. Finally, Figure~\ref{fig:relativeFreqDetection} shows the relative amount of true detections among all detections for $\ell \in \{0, 0.1, \ldots, 5\}$. To this end, we count in how many simulations the signal is \textit{correctly identified}, i.e., only the peak and no other change is detected. More precisely, we define the peak as \textit{detected} if there exists a $j$ such that $|\hat{\tau}_j - \tau_1| < m / \sr$ and $|\hat{\tau}_{j+1} - \tau_2| < m / \sr$ as a peak is shifted at most $m / \sr$ by the filter. If only one change but not a peak is within these boundaries (see the zooms in Figure \ref{fig:GramicidinJSMURF} for an illustration) we do not count it as a true detection, but also not as a \textit{false positive}, whereas all other changes are counted as \textit{false positives}. For the estimated change-point locations and the level we only consider cases where the peak is detected and report the mean squared error, the bias and the standard deviation. We also report trimmed versions for the estimated level, where we compute these error quantities only for all values between $0$ and $40$.

\begin{table}[ht]
\centering
 \caption{\footnotesize Performance of $\JULES$, $\TRANSIT$ and $\JSMURF$
 in idealizing a signal with an isolated peak having levels $l_0{\black = l_2}=40$,
 $l_1=20$, change-points $\tau_1=0.2$ and $\tau_2 = \tau_1 + \ell / \sr$,
 $\ell = 2,3,5$. The standard deviation of the error used in each simulation
 is $\sigma_0 = 1.4$. Results are based on $10\,000$ pseudo samples.}~\label{tab_singlePeak_detection}
\scalebox{0.75}{
  \begin{tabular}{l N N N N}
  \toprule[1.25pt]  
  Method & Length ($\ell$) & Correctly identified ($\%$) & Detected ($\%$) & False positive (Mean)\\
  \hline
  \\
 $\JULES$ & 2 & {\black   65.17 }  & {\black 65.17  }& {\black 0.0290 }\\ 
  $\TRANSIT$ & 2 & 0.01 & 96.02 & 19.9692 \\ 
  $\JSMURF$ & 2 & 0.00 & 0.00 & 0.2471 \\ 
  $\JULES$ & 3 &{\black 99.82 }&{\black 99.82 }&{\black 0.0004} \\ 
  $\TRANSIT$ & 3 & 0.02 & 74.65 & 19.8013 \\ 
  $\JSMURF$ & 3 & 0.00 & 0.00 & 0.1366 \\ 
   $\JULES$ & 5 &{\black 100.00 }& {\black 100.00 }&{\black 0.0000 }\\ 
  $\TRANSIT$ & 5 & 0.00 & 98.20 & 19.8484 \\ 
  $\JSMURF$ & 5 & 0.00 & 0.00 & 0.0000 \\
  {\black $\MDL$} & {\black 5} & {\black 0.02} & {\black 100.00} & {\black 96.8718}\\   
  \\
  \bottomrule[1.25pt]
  \end{tabular}
}
\end{table}

\begin{table}[ht]
\centering
 \caption{\footnotesize Performance of $\JULES$ and $\TRANSIT$
 in idealizing a signal with an isolated peak having levels $l_0{\black = l_2}=40$,
 $l_1=20$, change-points $\tau_1=0.2$ and $\tau_2 = \tau_1 + \ell/\sr$,
 $\ell = 2,3,5$. The standard deviation of the error used in each simulation
 is $\sigma_0 = 1.4$. Results are based on $10\,000$ pseudo samples. Results are given as multiples of the sampling rate $\sr = 10^4$.}~\label{tab_singlePeak_estimationLocation}
\scalebox{0.75}{
  \begin{tabular}{l N N N N N N N N}
  \toprule[1.25pt]  
    Method & Length ($\ell$) 
  & $\sr^2\MSE(\hat{\tau}_1)$ & $\sr\BIAS(\hat{\tau}_1)$ & $\sr\SD(\hat{\tau}_1)$ 
  & $\sr^2\MSE(\hat{\tau}_2)$ & $\sr\BIAS(\hat{\tau}_2)$ & $\sr\SD(\hat{\tau}_2)$\\ 
  \hline
  \\
 $\JULES$ & 2 & { 0.4022 }&{  -0.1047 }&{  0.6255 }&{  0.2677 }&{  0.0587 }&{  0.5141 }\\ 
  $\TRANSIT$ & 2 & 9.3427 & 2.6218 & 1.5713 & 24.1103 & 4.7920 & 1.0709\\ 
$\JULES$ & 3 &{  0.1170 }&{  0.0044 }&{  0.3420 }&{  0.1087 }&{  -0.0012 }&{  0.3297} \\ 
  $\TRANSIT$ & 3 & 13.0761 & 2.9386 & 2.1074 & 24.2032 & 4.7115 & 1.4162\\ 
 $\JULES$ & 5 &{  0.0670 }&{  -0.0026 }&{  0.2588 }&{  0.0669 }&{  0.0025 }&{  0.2587} \\ 
  $\TRANSIT$ & 5 & 15.4710 & 3.5622 & 1.6679 & 18.6559 & 4.0973 & 1.3670\\ 
  \\
  \bottomrule[1.25pt]
  \end{tabular}
}
\end{table}

\begin{table}[ht]
\centering
 \caption{\footnotesize Performance of $\JULES$ and $\TRANSIT$
 in idealizing a signal with an isolated peak having levels $l_0{\black = l_2}=40$,
 $l_1=20$, change-points $\tau_1=0.2$ and $\tau_2 = \tau_1 + \ell / \sr$,
 $\ell = 2,3,5$. The standard deviation of the error used in each simulation
 is $\sigma_0 = 1.4$. Results are based on $10\,000$ pseudo samples. Note that $\JULES$ overestimates the amplitude in very few cases (less than $0.2\%$) heavily while in all other simulations the estimation is very well. To illustrate this issue we also reported trimmed versions of all error quantities, where we only considered values between $0$ and $40$.}~\label{tab_singlePeak_estimationLevel}
\scalebox{0.75}{
  \begin{tabular}{l N N N N N N N N}
  \toprule[1.25pt]  
    Method & Length ($\ell$) 
  & $\MSE(\hat{\valmu}_1)$ & $\BIAS(\hat{\valmu}_1)$ & $\SD(\hat{\valmu}_1)$
  & $\MSE_{\operatorname{Trim}}(\hat{\valmu}_1)$ & $\BIAS_{\operatorname{Trim}}(\hat{\valmu}_1)$ & $\SD_{\operatorname{Trim}}(\hat{\valmu}_1)$\\ 
  \hline
  \\
  $\JULES$ & 2 & { 222978.6771 }&{  -83.6452 }&{  464.7745 }&{  32.5343 }&{  -0.6010 }&{  5.6726} \\  
 $\TRANSIT$ & 2 & 112.7163 & 10.1733 & 3.0367 & 99.3660 & 9.7529 & 2.0609 \\ 
 $\JULES$ & 3 &{  552.8490 }&{  -1.0170 }&{  23.4919 }&{  12.3488 }&{  -0.7219 }&{  3.4393 }\\ 
  $\TRANSIT$ & 3 & 69.6075 & 7.0327 & 4.4891 & 47.2984 & 6.2367 & 2.8988 \\ 
 $\JULES$ & 5 &{  2.7763 }&{  -0.1081 }&{  1.6628 }&{  2.7763 }&{  -0.1081 }&{  1.6628} \\ 
  $\TRANSIT$ & 5 & 26.8554 & 2.0811 & 4.7462 & 11.5351 & 1.4210 & 3.0849 \\ 
  \\
  \bottomrule[1.25pt]
  \end{tabular}
}
\end{table}

We found that $\JULES$ detection power increases with the length $\ell$ in range between $0.8$ to $2.4$ times the sampling rate and is (almost) one for larger lengths while at the same time it provides almost no false positives, resulting in a relative amount of true detections of almost one for lengths larger than one quarter of the filter lengths. In comparison, $\TRANSIT$ has a slightly larger detection power for very small lengths, but detects much more false positives, resulting in almost no correctly identified signals. Moreover, its detection power fluctuates for larger lengths in an uncontrollable way. $\JSMURF$ is not able to detect such small filtered peaks as it does not take into account the corresponding scales, but also does not detect false positives. {\black We also briefly investigated $\MDL$ \cite{gnanasambandam2017unsupervised}, see Table \ref{tab_singlePeak_detection}. To decrease the amount of false positives the authors suggested to assume a minimal length for the events. Although this might be problematic for real data applications with short events, we used for the simulations with $l=5$ the assumption of that length as prior information. But even then the number of events is heavily overestimated. More precisely, on average $98.87$ change-points are discovered instead of the two change-points of the peak. Note, that the simulated observations are filtered by a 4-pole lowpass Bessel filter with normalized cutoff frequency of $0.1$ and as noted by \cite{gnanasambandam2017unsupervised} their approach works better when recordings are unfiltered or only slightly filtered.}\\
$\JULES$ idealizes almost all detected peaks with high accuracy which increases with the length of the peak. {\black Only in very rare cases (less than $0.2\%$) and only if the peak is short the amplitude is heavily overestimated}, resulting in large mean squared error, bias and standard deviation, as the corresponding trimmed values are much smaller. We still found that the convolved idealization fits the observations always very well, see Figure~\ref{fig:singlePeak_a} for an illustration. We will also see in Section \ref{sec:hmm} that $\JULES$ is nevertheless able to identify two levels close to each other. In comparison, Figure~\ref{fig:singlePeak_b} and the biases in Tables \ref{tab_singlePeak_estimationLocation} and \ref{tab_singlePeak_estimationLevel} show that $\TRANSIT$ miss-estimates the peak systematically, in particular the amplitude is underestimated if the peak is smoothed by the filter. The major difficulty of the devonvolution problem is that convolutions of signals with larger amplitude but smaller length can look very similar to the convolution of a signal with smaller amplitude but larger length. In fact, it is possible to show that the change-points locations $\tau_i$ can by no method be estimated better than at $1/\sqrt{n}$ rate (instead of the sampling rate $1/n$ without convolution), see \cite{Boysen.etal.09} for a similar setting and also \cite{frick2014asymptotic, goldenshluger2006optimal} for further theoretical results of estimation a peak from filtered data. Note that the filter is rather short ranged with a filter length of eleven but the signal to noise ratio is rather small with $20 / 1.4 \approx 14.29$, since gA has a small conductance in comparison to other proteins, and the lengths are very short with two to five observations.\\
Similar results were obtained in additional simulation studies (not included) with different amplitude and noise level. Also a simulation (not displayed) where we shift the two changes by $0.1 / \sr, 0.25 / \sr, 0.5 / \sr, 0.75 / \sr$ and $0.9 / \sr$ leads to almost identical results for $\JULES$. This confirms the ability of $\JULES$ to identify correctly the location (and amplitude) of changes between the sampling points.

\subsection{Separation of two consecutive peaks}~\label{sec:consecutivePeaks}
\begin{figure}[htb]
  \begin{subfigure}[t]{0.32\linewidth}
    \centering
    \includegraphics[width=\linewidth]{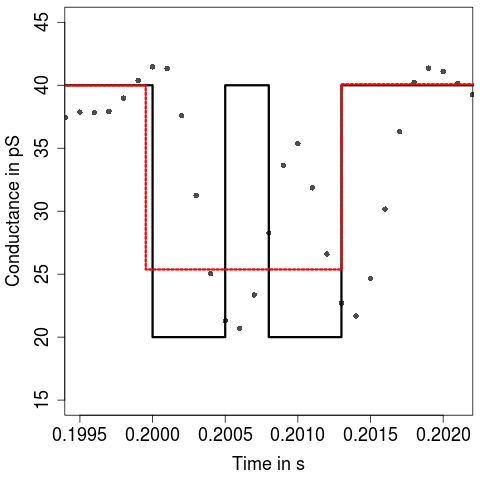} 
    \caption{\footnotesize{No separation in the detection step, $d = 3$}}
    \label{fig:twoPeaks_b} 
  \end{subfigure}
  \begin{subfigure}[t]{0.32\linewidth}
    \centering
    \includegraphics[width=\linewidth]{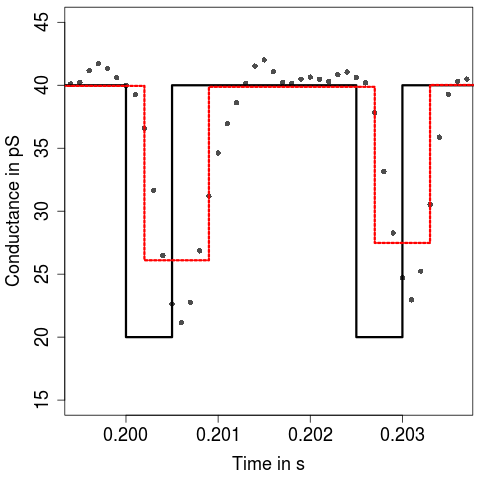} 
    \caption{\footnotesize{No separation in the deconvolution, $d = 20$}}
    \label{fig:twoPeaks_c} 
  \end{subfigure} 
  \begin{subfigure}[t]{0.32\linewidth}
    \centering
    \includegraphics[width=\linewidth]{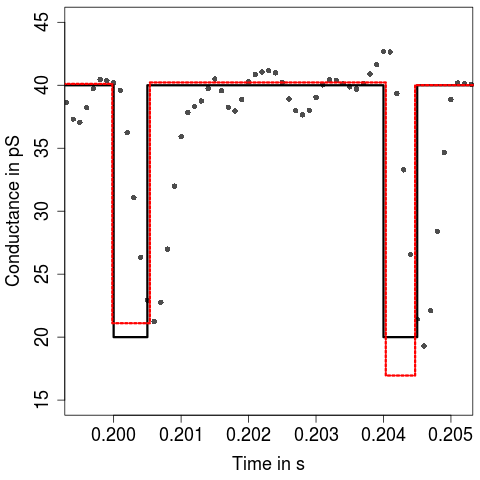} 
    \caption{\footnotesize{Perfect separation, $d = 35$}}
    \label{fig:twoPeaks_d} 
  \end{subfigure}
  \caption{\footnotesize{Data $y_i = F \ast (f(i/n) + \sigma_0 \epsilon_i)$ (grey points), where $\sigma_0=1.4$, $\epsilon_i$ is gaussian white noise and the signal $f$  has two consecutive peaks comprised of the levels $l_0{\black = l_2 = l_4} = 40$, $l_1{\black = l_3} = 20$ and change-points $\tau_1 = 2000 / \sr$, $\tau_2 = \tau_1 + 5/\sr$, $\tau_3 = \tau_2 + d$ and $\tau_4 = \tau_3 + 5/\sr$. True signal ({\black \solidrule}) and $\JULES$ idealization ({\ttlred  \solidrule}). Idealization coincides with reconstruction from the detection step if separation fails in the deconvolution step.}}
  \label{fig:twoPeaks} 
\end{figure}

In this simulation study we are interested in the performance of $\JULES$ in separating two consecutive peaks as a function of $d$, the distance between them. Separation is necessary in the detection step, i.e., two peaks have to be detected, but also in the deconvolution step as we are only able to deconvolve both peaks individually if they are isolated. To this end, we consider a signal $f$ with change-points $\tau_1 = 2\,000 / \sr$, $\tau_2 = \tau_1 + 5/\sr$, $\tau_3 = \tau_2 + d$ and $\tau_4 = \tau_3 + 5/\sr$, with $\tau_0 = 0$ and $\tau_{\operatorname{end}} = 4\,000/\sr$ and levels $l_0=l_2=l_4=\SI{40}{\pico\siemens}$ and $l_1=l_3=\SI{20}{\pico\siemens}$. Here we fixed the length of the consecutive peaks at $5/\sr$, since want to focus on separation only and we found in Section~\ref{sec:singlePeak} that peaks of this size can be detected with (almost) probability one. In this section we focus only on $\JULES$ as separation signifies an issue for this method.\\
We identified three outcomes for $\JULES$, all illustrated in Figure~\ref{fig:twoPeaks}. First of all, perfect separation, i.e., the multiresolution reconstruction of $\JULES$ identifies the two peaks (4 change-points) and the local deconvolution yields idealizations for the four levels. Secondly, separation fails in the detection step, i.e., the multiresolution reconstruction recognizes only 2 change-points and identifies one peak whose level can be further deconvolved. Finally, separation fails in the deconvolution step, i.e., $\JULES$ identifies two peaks but the distance is so small that the deconvolution method cannot separate them, in other words, no long segment is in between.

\begin{figure}[htb]
  \centering
    \includegraphics[width=.98\textwidth, keepaspectratio]{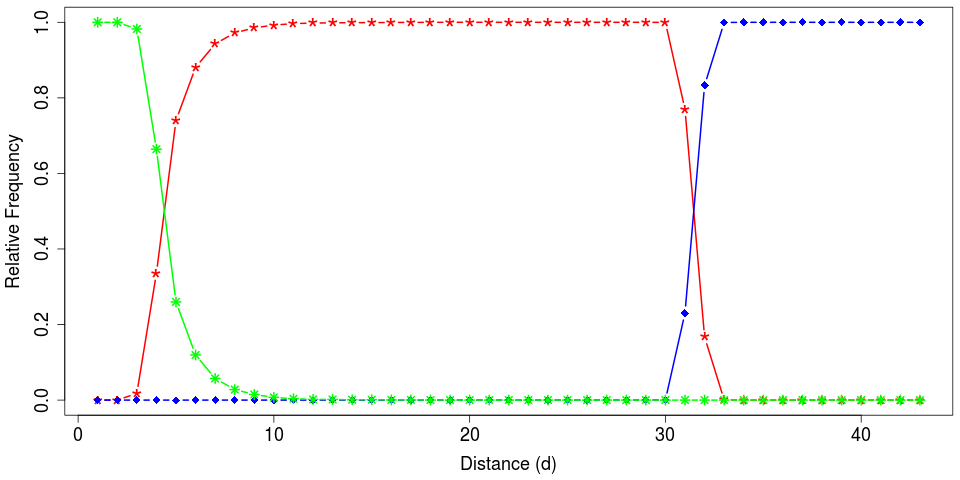}
    \caption{\footnotesize{$\JULES$'s behavior in idealizing two consecutive peaks separated by distance $d$. See Figure~\ref{fig:twoPeaks} for specifications of signal and noise used in this case. $\JULES$'s frequency of no separation in the detection step ({\green \dotsrule}). $\JULES$'s frequency of successful detection, but no separation in the deconvolution step ({\mylightred \dashedrule}). $\JULES$'s frequency of successful detection and deconvolution ({\blue \solidrule}). Results are based on $10\,000$ simulations for each value of $d$.}}
    \label{fig:peakSeparation} 
\end{figure}

Figure~\ref{fig:peakSeparation} shows the frequency at which each scenario occurred as a function of $d$, the distance between the two peaks, in $10\,000$ simulations for each value of $d = \{1, 2, \ldots, 43\}$. We found that the detection of both peaks fails potentially when $d \leq 12$, but the frequency decreases rapidly with $d$. For $d \in [4, 31]$ detection is often possible, but not the separation in the deconvolution, whereas for $d > 33$ this scenario is no longer observed. Finally, separation of the two peaks is possible with high probability as soon as the distance between them is at least $32 / \sr$, roughly three times the filter lengths. This is equal to the minimal distance of a long segment $10$ plus the two shifts on the left and right side of the segment by $m = 11$ to take the filter into account. This corresponds to $\SI{3.2}{\milli\second}$ for the gA traces, whereas the estimated average distance is with $\SI{1 / 3.28}{\second}\approx \SI{0.3}{\second}$ (see Section \ref{sec:analysis}) much larger. {\black Moreover, we found that the distance required to detect two separated peaks is only roughly three times the length that is required to detect a single, isolated peak (confer Figure \ref{fig:relativeFreqDetection}). This shows that $\JULES$ separates peaks well with respect to detection.}

\subsection{Hidden Markov model}~\label{sec:hmm}
\begin{figure}[!htp]
\centering
\begin{subfigure}{0.32\textwidth}
\includegraphics[width = \textwidth]{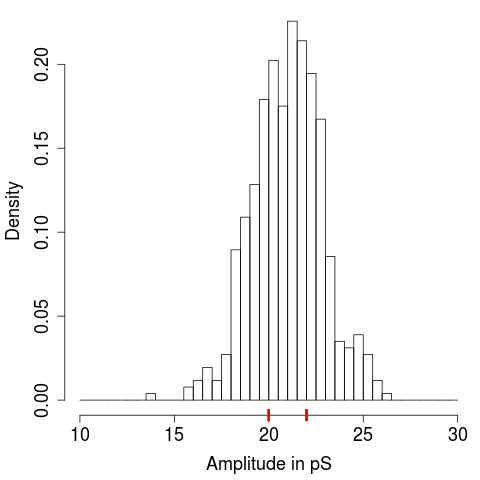}
\subcaption{$\JULES$, $\Delta = 2$}
\label{subfig:A:simampjules22}
\end{subfigure}
\begin{subfigure}{0.32\textwidth}
\includegraphics[width = \textwidth]{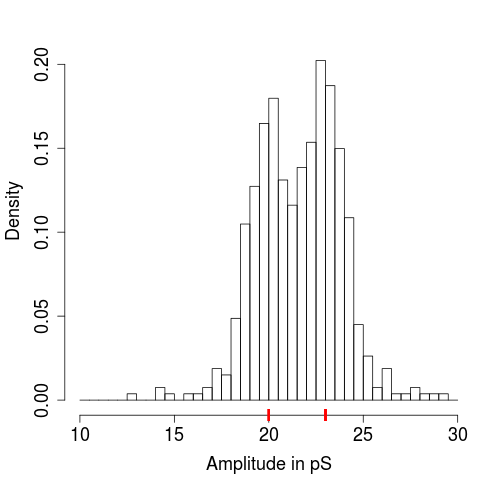}
\subcaption{$\JULES$, $\Delta = 3$}
\label{subfig:A:simampjules23}
\end{subfigure}
\begin{subfigure}{0.32\textwidth}
\includegraphics[width = \textwidth]{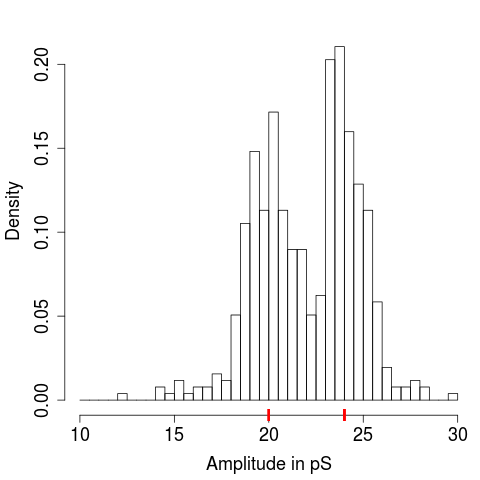}
\subcaption{$\JULES$, $\Delta = 4$}
\label{subfig:A:simampjules24}
\end{subfigure}\\
\begin{subfigure}{0.32\textwidth}
\includegraphics[width = \textwidth]{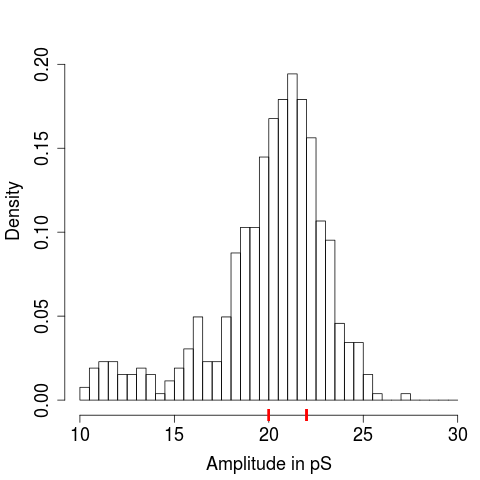}
\subcaption{$\TRANSIT$, $\Delta = 2$}
\label{subfig:B:simamptransit22}
\end{subfigure}
\begin{subfigure}{0.32\textwidth}
\includegraphics[width = \textwidth]{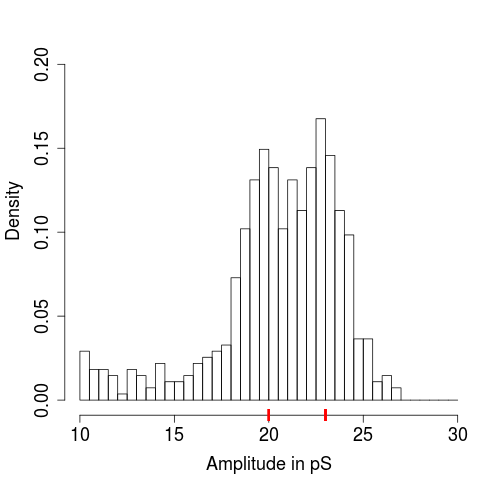}
\subcaption{$\TRANSIT$, $\Delta = 3$}
\label{subfig:B:simamptransit23}
\end{subfigure}
\begin{subfigure}{0.32\textwidth}
\includegraphics[width = \textwidth]{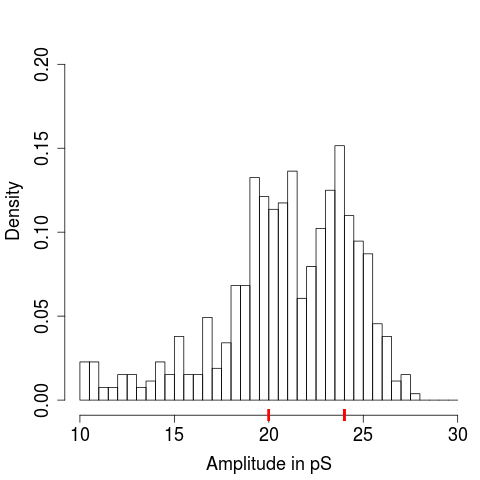}
\subcaption{$\TRANSIT$, $\Delta = 4$}
\label{subfig:B:simamptransit24}
\end{subfigure}
\caption{Histograms of the amplitudes of idealized events for various $\Delta$ and dwell time below $\SI{2.6}{\milli\second}$. Red tick marks are the true amplitudes (both with equal probability).}
\label{fig:hmmlevel}
\end{figure}

In this section we simulate data from a three state Hidden Markov model. None of the considered methods require a Hidden Markov assumption, but as such a model is often assumed for ion channel recordings, it is instructive to investigate these methods in such a scenario. More precisely, we simulate one open state with $\SI{40}{\pico\siemens}$ and two closed states with $\SI{20}{\pico\siemens}$ and $20+\Delta$\SI{}{\pico\siemens}, i.e., the amplitudes of the two flickering states ($\SI{20}{\pico\siemens}$ and $20-\Delta$\SI{}{\pico\siemens}) differ by $\Delta$. The dwell time in the open state is exponentially distributed with rate $2.5$ and the channel switches to both closed states equally likely. For both closed states the channel reopens quickly with rate $800$ (\SI{0.8}{\kilo\hertz}). We generate five traces with $600\,000$ observations, each.\\
In the following we analyze these data in a similar fashion as we will do it for the gramicidin A traces in Section \ref{sec:analysis}. In accordance with the definition of a short segment in Section \ref{sec:deconvolution} we define a closing event as a flickering event if its dwell time is smaller than or equal to $\SI{2.6}{\milli\second}$, but results are qualitative the same (amplitude histograms are even quantitative very similar) if we increase or decrease this threshold within a reasonable range. We also exclude events with a dwell time smaller than $\SI{0.24}{\milli\second}$ as we cannot detect such events reliably and they would disturb the analysis. We do not include results for $\JSMURF$, since it is not able to detect such small flickering events, c.f. Figure \ref{fig:relativeFreqDetection}. Figure \ref{fig:hmmlevel} shows histograms of the idealized amplitudes of flickering events with a small bin width of $0.5$ (visually chosen) to already see small indications of two different levels. 

\begin{figure}[!htp]
\centering
\begin{subfigure}{0.49\textwidth}
\includegraphics[width = \textwidth]{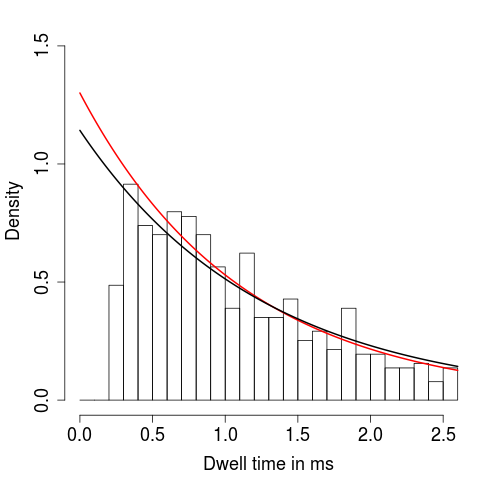}
\subcaption{$\JULES$}
\label{subfig:A:hmmdwelljules}
\end{subfigure}
\begin{subfigure}{0.49\textwidth}
\includegraphics[width = \textwidth]{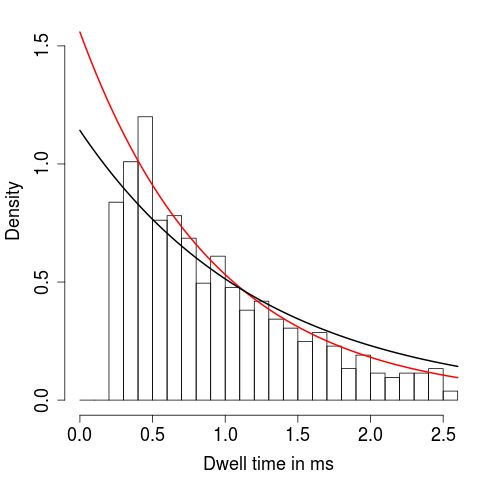}
\subcaption{$\TRANSIT$}
\label{subfig:B:hmmdwelltransit}
\end{subfigure}
\caption{Histograms of the dwell time in the closed state for $\Delta=2$ for all closing events with amplitude between $10$ and $\SI{30}{\pico\siemens}$ together with the true exponential distribution with parameter $0.8$ ({\black \solidrule}) and exponential fits ({\ttlred  \solidrule}). We rescaled all lines such that the area under them are standardized to one to make them comparable to the histograms.}
\label{fig:hmmdwell}
\end{figure}

In Table \ref{tab_singlePeak_estimationLevel} we found that $\JULES$ level estimate has a large mean squared error, mainly due to heavy overestimations in rare cases. Nevertheless, we see in this simulation that $\JULES$ idealizes and separates the two amplitudes very well. For $\Delta = 2$ a tiny indication of two different amplitudes can be seen, for $\Delta = 3$ the two peaks are distinct and for larger differences the two different states can be clearly detected. In comparison, $\TRANSIT$ finds some smaller amplitudes and hence the separation is slightly less clear, but still possible for $\Delta \geq 3$. We already saw in Figure \ref{fig:singlePeak} and Table \ref{tab_singlePeak_estimationLevel} that $\TRANSIT$ estimates a too small amplitude if a short peak is smoothed by the filter.\\
Figure \ref{fig:hmmdwell} shows the dwell times in the closed state and an exponential fit. More precisely, we only consider events with a dwell time such that we can detect the events reliably, i.e., events with dwell time between $0.24$ and $\SI{2.6}{\milli\second}$ and estimate the rate by maximizing the likelihood of these events assuming an exponential distribution for the dwell times. Note that the results for the dwell time and also for the later analyzed distances are similar for all considered $\Delta$. We show results exemplary for $\Delta = 2$.

\begin{figure}[!htp]
\centering
\begin{subfigure}{0.49\textwidth}
\includegraphics[width = \textwidth]{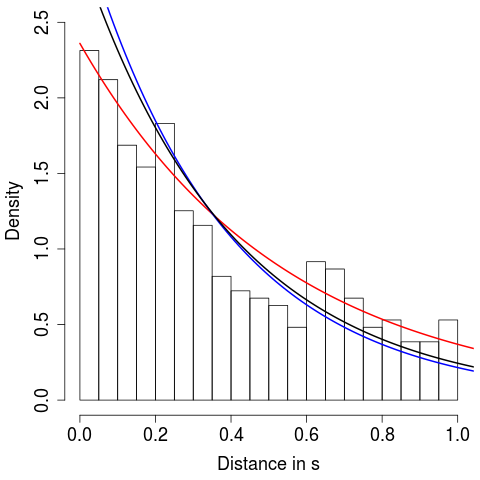}
\subcaption{$\JULES$}
\label{subfig:A:hmmdistancejules}
\end{subfigure}
\begin{subfigure}{0.49\textwidth}
\includegraphics[width = \textwidth]{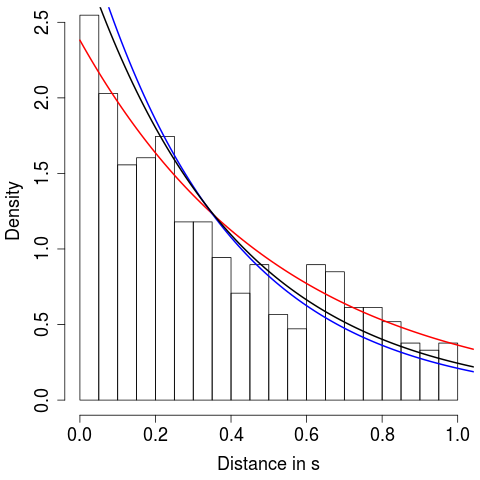}
\subcaption{$\TRANSIT$}
\label{subfig:B:hmmdistancetranit}
\end{subfigure}
\caption{Histograms of the distance between two flickering events, i.e., events with amplitude between $10$ and $\SI{30}{\pico\siemens}$ and dwell time below $\SI{2.6}{\milli\second}$, for $\Delta=2$ together with the true exponential distribution with parameter $2.5$ ({\black  \solidrule}), exponential fits ({\ttlred  \solidrule}) and the exponential fits corrected for missed events ({\blue  \solidrule}). We rescaled all lines such that the area under them are standardized to one to make them comparable to the histograms.}
\label{fig:hmmdistance}
\end{figure}

Apart from the fact that both methods miss extremely short events (\SI{< 0.3}{\milli\second}) which coincides with Figure \ref{fig:relativeFreqDetection}, $\JULES$ confirms an exponential distribution and estimates with $0.9$ roughly the right parameter. $\TRANSIT$ detects additional spurious events with lengths between $0.4$ and $\SI{0.5}{\milli\second}$ and hence overestimates the rate with $1.08$ slightly more. A similar behavior could be observed in Table \ref{tab_singlePeak_estimationLocation}, too.\\
Figure \ref{fig:hmmdistance} shows the distance between two flickering events which coincides with the dwell time in the open state if other closing events are considered as spurious events. We include events with a distance between $0.032$ and $\SI{1}{\second}$ as Figure $\ref{fig:peakSeparation}$ shows that $\JULES$ is not able to separate peaks with a smaller distance. However, such a fit is not enough as we miss events. Hence, a precise recovery of the rates requires to recalculate the distribution by taking into account missed events, see for instance \cite{hawkes.etal.90, colquhoun1996joint, qin1996estimating, epstein2016bayesian}. In this case simply dividing the estimated rate by the probability that an event has dwell time between $0.24$ and $\SI{2.6}{\milli\second}$ assuming an exponential distribution with the fitted rate appears to be appropriate.

For both methods the distribution of the distances seems to be exponential, however, due to the missing events the rate $2.5$ cannot be recovered by a simple exponential fit as it is the case for the dwell times. For $\JULES$ we find a rate of $1.86$ and with $\TRANSIT$ of $1.88$. Including the missed events correction leads with $2.69$ and $2.72$, respectively, to satisfactory results.

{\black
\subsection{Robustness}\label{sec:robustness}

While filtered white noise is already a good approximation for the noise occurring in the data analyzed here, see Section \ref{sec:analysis}, additional high frequent $f^2$ (violet) and long tailed $1/f$ (pink) noise components have been observed in patch clamp recordings as well, for a more detailed discussion see \cite{Neher.Sakmann.76, venkataramanan1998identification, levis1993use} and the references therein. Moreover, heterogeneous noise can be caused by open channel noise \cite{venkataramanan1998identification}, i.e., the noise level can be larger when the conductance is larger. Therefore, we focus in this section on analyzing how robust $\JULES$ is against such noise. To simplify the data generation we use a discrete convolution for the filter. For the violet noise we use as suggested by \cite{venkataramanan1998} a moving average process with coefficients $0.8$ and $-0.6$. For the pink noise we use the algorithm available on \textit{https://github.com/Stenzel/newshadeofpink}. In both cases we scale the noise such that each single observation has standard deviation $1.4$ (to make it comparable to the simulations in the previous sections) and such that half of it results from the white noise and half of it from the additional noise component. For the heterogeneous noise case we simulate standard deviation $1.4$ for the background noise and $2.8$ for the events by oversampling the recordings by a factor of $100$.

\begin{table}[ht]
\centering
 \caption{\footnotesize \black Robustness of $\JULES$ against additional noise components
 in idealizing a signal with an isolated peak having levels $l_0=l_2=40$,
 $l_1=20$, change-points $\tau_1=0.2$ and $\tau_2 = \tau_1 + \ell / \sr$,
 $\ell = 2,3,5$. The standard deviation of the error used in each simulation
 is $\sigma_0 = 1.4$. Results are based on $10\,000$ pseudo samples.}~\label{tab_singlePeak_robustness}
\scalebox{0.75}{\black
  \begin{tabular}{l N N N N}
  \toprule[1.25pt]  
  Method & Length ($\ell$) & Correctly identified ($\%$) & Detected ($\%$) & False positive (Mean)\\
  \hline
  \\
  White noise & 2 & 64.42 & 64.42 & 0.0333 \\ 
  $f^2$ noise & 2 & 65.56 & 65.56 & 0.0219 \\ 
  $1/f$ noise & 2 & 78.43 & 99.93 & 0.2306 \\
  Heterogeneous noise & 2 & 59.05 & 59.05 & 0.0214 \\ 
  White noise & 3 & 99.82 & 99.82 & 0.0005 \\ 
  $f^2$ noise & 3 & 99.88 & 99.88 & 0.0000 \\ 
  $1/f$ noise & 3 & 78.02 & 100.00 & 0.2379 \\
  Heterogeneous noise & 3 & 95.23 & 95.23 & 0.0023 \\
  White noise & 5 & 100.00 & 100.00 & 0.0000 \\ 
  $f^2$ noise & 5 & 100.00 & 100.00 & 0.0000 \\ 
  $1/f$ noise & 5 & 73.60 & 100.00 & 0.2911 \\
  Heterogeneous noise & 5 & 99.95 & 99.95 & 0.0000 \\  
  \\
  \bottomrule[1.25pt]
  \end{tabular}
}
\end{table}

From Table \ref{tab_singlePeak_robustness} we conclude that $\JULES$ is very robust against the additional $f^2$ and heterogeneous noise but influenced by $1/f$ noise. At presence of the latter noise, the global standard deviation is underestimated by \eqref{eq:estsigma0} which leads to a larger detection power, but at the price of additional false positives. Note, that false positives are caused by the underestimated standard deviation but also by the long range dependency itself. However, these false positives have a small amplitude and therefore do not influence the analysis severely or can be removed by postfiltering. At presence of heterogeneous noise the detection power is slightly decreased due to the larger variance. However, we stress that $\JULES$ is only robust against heterogeneous noise as long as only short events occur. If we would have also some longer events the non-stationarity caused by open channel noise would be a severe issue. Parameter estimation (not displayed) is slightly worse at presence of $f^2$ or heterogeneous noise, but not affected by presence of $1/f$ noise.
}

\section{Data analysis}\label{sec:analysis}
\subsection{Measurements}

Gramicidin A was obtained from Sigma-Aldrich (Schnelldorf, Germany) and used without further purification. 1,2-Diphytanoyl-\textit{sn}-glycero-3-phosphocholine (Avanti Polar Lipids, Alabaster, USA) and cholesterol (Sigma) were dissolved in chloroform and mixed in a molar ratio of 9:1. Giant-unilamellar vesicles (GUVs) containing the peptide (peptide to lipid ratio of 1:10000 up to 1:1000) were obtained by electroformation \cite{Angelova:1986}. Voltage clamp experiments of solvent-free lipid bilayers were performed using the Port-a-Patch (Nanion Technologies, Munich, Germany) in buffer ($\SI{1}{\Molar}$ KCl, $\SI{10}{\milli\Molar}$ HEPES, pH 7.4). Electrically insulating membranes with resistances $\SI{>1}{\giga\ohm}$ were obtained by spreading GUVs on the aperture of a glass chip applying a slight negative pressure. A DC potential of $+\SI{100}{\milli\volt}$ was applied and data was recorded at a sampling rate of $\SI{10}{\kilo\hertz}$ using a $\SI{1}{\kilo\hertz}$ 4-pole Bessel filter.


\subsection{Idealization}
Idealizations are obtained by $\JULES$, $\TRANSIT$ and $\JSMURF$ with parameter choices as in Section \ref{sec:parameterchoices}. The recorded observations in Figure \ref{fig:GramicidinData} show gating events between two states on various time scales, but also several noise effects like outliers or varying conductivity, compare for instance the conductivity from $10$ to $25$ seconds and from $55$ to $60$ seconds. Note that such effects raise substantial difficulties for methods that assumes a Hidden Markov model or similar models. Contrary, $\JULES$ (Figure \ref{fig:GramicidinJULES}) provides a reasonable idealization covering all major features of the data and some smaller fluctuations. In comparison, $\TRANSIT$ (Figure \ref{fig:GramicidinTRANSIT}) is able to detect such events too but at the same hand detects far too many false positives. $\JSMURF$ (Figure \ref{fig:GramicidinJSMURF}) works well on larger time scales but misses flickering events below the filter length. The zooms into single peaks (Figures \ref{fig:GramicidinJULES}-\ref{fig:GramicidinTRANSIT} lower panels) demonstrates that $\JULES$ fits the observation well, whereas $\TRANSIT$ estimates a too small amplitude due to the smoothing of the filter and $\JSMURF$ misses the peaks.

\subsection{Analysis of flickering dynamics}

\begin{figure}[!htp]
\centering
\begin{subfigure}{0.49\textwidth}
\includegraphics[width = \textwidth]{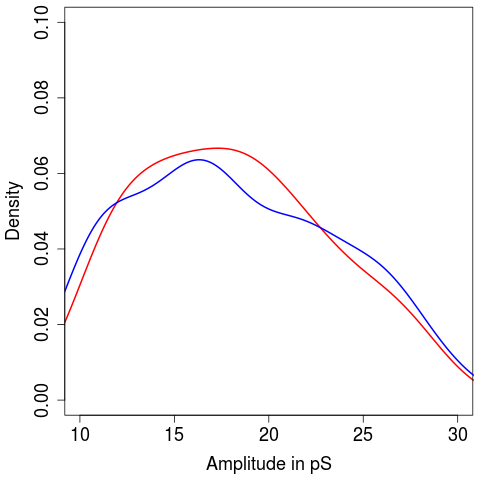}
\subcaption{$\JULES$}
\label{subfig:A:histlevelpeaks}
\end{subfigure}
\begin{subfigure}{0.49\textwidth}
\includegraphics[width = \textwidth]{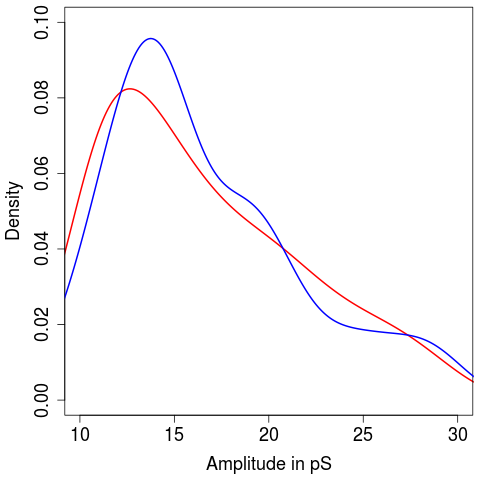}
\subcaption{$\TRANSIT$}
\label{subfig:B:histvaluejumpssingletrauncated}
\end{subfigure}
\caption{Kernel density estimates with bandwidth $2$ of the amplitudes of events with dwell time below $\SI{2.6}{\milli\second}$ ({\ttlred  \solidrule}) and above $\SI{10}{\milli\second}$ ({\blue  \solidrule}).}
\label{fig:histlevel}
\end{figure}

In this section we analyze the flickering dynamics of gramicidin A based on the obtained idealizations in a similar fashion as the simulated data in Section \ref{sec:hmm}. Note that flickering does not occur in all data traces, for this analysis we focus on five traces, each with $600\,000$ data points, which show significant flickering. As before we define a closing event as a flickering event if its dwell time is smaller than or equal to $\SI{2.6}{\milli\second}$. For comparison we also study briefly the amplitude of the slow gating (closing events between segments with dwell time longer than $\SI{10}{\milli\second}$). We focus on events with an amplitude larger than $\SI{\geq 10}{\pico\siemens}$ as events with a smaller amplitude are more likely base line fluctuations than channel gating. Base line fluctuations are for instance caused by small defects in the membrane, which is unavoidable in
the recordings. There might be also periodic oscillations, resulting from the electronic or from building vibrations (although damped). Figure \ref{fig:histlevel} compare the idealized amplitudes of flickering events between $10$ and $\SI{30}{\pico\siemens}$ ($307$ events) with the amplitudes of slow gating events ($44$ events) of the same magnitude by kernel density estimates. The latter are performed by the \texttt{R} function \texttt{bkde} in the package \texttt{KernSmooth} \cite{wand.jones.94} with bandwidth $2$ (visually chosen).

\begin{figure}[!htp]
\centering
\begin{subfigure}{0.49\textwidth}
\includegraphics[width = \textwidth]{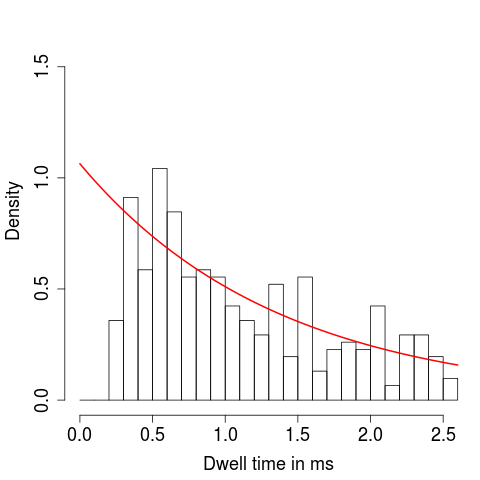}
\subcaption{$\JULES$}
\label{subfig:A:dwelljules}
\end{subfigure}
\begin{subfigure}{0.49\textwidth}
\includegraphics[width = \textwidth]{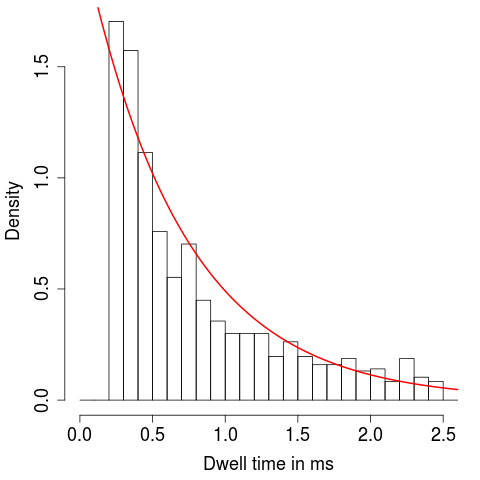}
\subcaption{$\TRANSIT$}
\label{subfig:B:dwelltransit}
\end{subfigure}
\caption{Histograms of the dwell time in the closed state for all closing events with amplitude between $10$ and $\SI{30}{\pico\siemens}$ together with exponential fits ({\ttlred  \solidrule}) which are rescaled such that the area under them are standardized to one to make them comparable to the histograms.}
\label{fig:dwell}
\end{figure}

\begin{figure}[!htp]
\centering
\begin{subfigure}{0.49\textwidth}
\includegraphics[width = \textwidth]{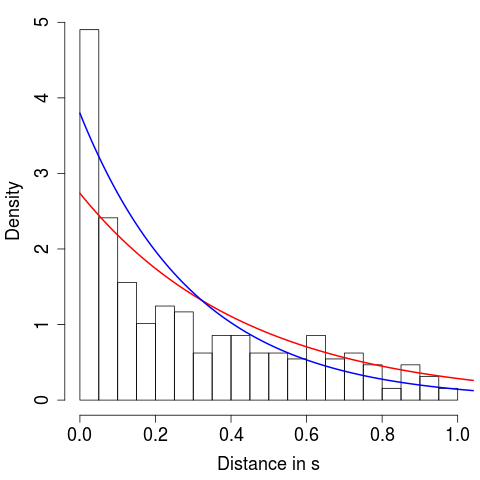}
\subcaption{$\JULES$}
\label{subfig:A:distancejules}
\end{subfigure}
\begin{subfigure}{0.49\textwidth}
\includegraphics[width = \textwidth]{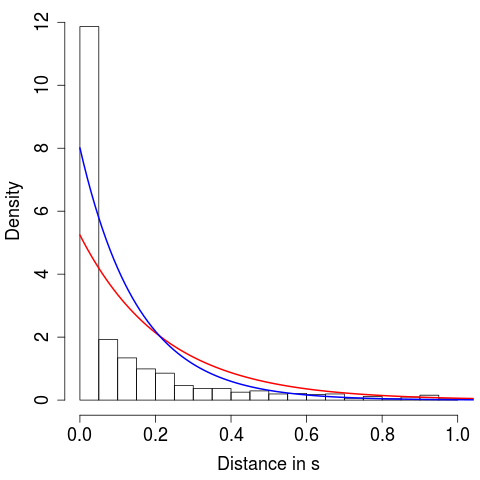}
\subcaption{$\TRANSIT$}
\label{subfig:B:distancetranit}
\end{subfigure}
\caption{Histograms of the distance between two flickering events, i.e., events with amplitude between $10$ and $\SI{30}{\pico\siemens}$ and dwell time below $\SI{2.6}{\milli\second}$ together with exponential fits ({\ttlred  \solidrule}) and the fits corrected for missed events ({\blue  \solidrule}). All lines are rescaled such that the area under them are standardized to one to make them comparable to the histograms.}
\label{fig:distance}
\end{figure}

We see no distinct peak as in the simulations in Figure \ref{fig:hmmlevel}. {\black Note that this also true for histograms instead of kernel density fits, but kernel density fits allow a better comparison of flickering and the gating on larger time scales.} Hence, either multiple levels are underlying or more likely additional errors occur. Recall for instance the conductivity fluctuations in Figure \ref{fig:GramicidinData}. More importantly, we found that the flickering events are full-sized (have the same amplitude as the slow gating events). In comparison, $\TRANSIT$ suggests a smaller amplitude which can be explained by the fact that $\TRANSIT$ ignores the smoothing effect by the filter. We already saw this effect in the simulations in Section \ref{sec:simulations}, see Figures \ref{fig:singlePeak} and \ref{fig:hmmlevel} as well as Table \ref{tab_singlePeak_estimationLevel}. We do not show results for $\JSMURF$, since only very few flickering events have been detected, also this was observed before in Table \ref{tab_singlePeak_detection} and Figure \ref{fig:relativeFreqDetection} in the simulations.\\
For the same reason we exclude $\JSMURF$ from the following analysis of the dwell times and distances between two flickering events. Figure \ref{fig:dwell} shows the dwell times in the closed state and Figure \ref{fig:distance} shows the distance between two flickering events. The latter coincides with the dwell time in the open state if other events, in particular events with smaller amplitude, are considered as noise and not as gating.

Although the lipid system is totally different, our results are of a similar order than in \cite{Ring1986brief, Armstrong2002origin}. Flickering events occur on average roughly every second and are only around a millisecond long. We miss extremely short events (\SI{< 0.24}{\milli\second}) in accordance with the simulations, see Figures \ref{fig:relativeFreqDetection} and \ref{fig:hmmdwell}. Up to this, we can confirm an exponential distribution with rate $0.73$ (in $\SI{}{\milli\second}$) for the dwell times. In comparison, $\TRANSIT$ detects additional short events which, however, seem to be artifacts, see Figure \ref{fig:GramicidinTRANSIT} and Table \ref{tab_singlePeak_detection} in the simulations, resulting in a rate of $1.46$.\\
{\black From Figure \ref{fig:distance} we draw that too many short distances are observed for a good exponential fit. Since such an effect was not observed in the simulations, cf. Figure \ref{fig:hmmdistance}, we speculate that this is caused by artifacts. Using also here a missed event correction leads to an estimated rate of \SI{3.28}{\hertz} based on the idealization by $\JULES$. In contrast, when we use $\TRANSIT$ we estimate a higher rate of \SI{6.5}{\hertz} due to its many additional findings which are likely to be wrong, confer the number of false positives in Table \ref{tab_singlePeak_detection}. We stress that we cannot exclude that a mixture of two (or more) exponential distributions is underlying which could be for instance caused by multiple underlying biochemical processes provoking flickering events, but a more detailed analysis would require more events or less artifacts.}

\section{Discussion {\black and Outlook}}\label{sec:discussion}

We proposed a new method $\JULES$ for idealizing ion channel recordings, particularly suited for detecting and idealizing very short flickering events, while at the same time it idealizes events on intermediate and large scales. Detection and idealization of flickering is a difficult task due to pre-lowpass filtering incorporated in the amplifier. We confirmed its ability to detect and idealize such peaks in simulations and in an application on gramicidin A traces. One limitation is that two consecutive peaks have to be separated by enough observations (roughly three times the filter length in the given setting, see Figure \ref{fig:peakSeparation}) to enable a quick local deconvolution as in the paper. This can be refined at computational expenses. For the given data this does not seem to be a significant limitation as the estimated average distance is with $\SI{0.3}{\second}$ much larger than the needed distance of $\SI{3.3}{\milli\second}$.\\
Another improvement could be done in the detection step. {\black The detection power can be increased by using the interval system of all intervals instead of the system with all intervals containing a dyadic number of observations. However, this increases the computational complexity from $\mathcal{O}(n\log(n))$ to $\mathcal{O}(n^2)$. In the same spirit, the multiscale statistic $M$ in \eqref{eq:multistats} could be based on local likelihood ratio test statistics instead on partial sums. However, the likelihood ratio test statistic is computationally more expensive as it involves inverting the corresponding covariance matrices on each interval considered in \eqref{eq:multistats} and an empirical study (not displayed) has shown that reconstructions based on local likelihood ratio tests does only slightly improve detection power.\\} 
The current version of $\JULES$ requires an explicit knowledge of the filter which is usually known. The \textit{clampSeg} package has an implementation for the most commonly used Bessel filter (with an arbitrary number of poles and cutoff frequency). If the filter is not known or if no implementation is available, the covariance could be estimated purely data driven by difference based methods \cite{Munk.Tecuapetla.15} from the recordings as well. Then, from the estimated covariance a discrete kernel for the filter could be computed and used for the deconvolution instead of the time-continuous kernel.\\
{\black While we assume a non-stationary signal, we require a stationary noise. We found in Section \ref{sec:robustness} that $\JULES$ is rather robust against heterogeneous noise as long as all events are very short. In general, $\JULES$ assumes a global standard deviation and will be influenced by changes of the standard deviation during the recording, as e.g. caused by open channel noise. We are currently working on a follow up paper which will be able to deal with heterogeneous noise to overcome this limitation. We expect such a method also to be robust against changes of the correlation structure.}\\ 
Besides these possible improvements, we will continue with using $\JULES$ to analyze further recordings with high biological relevance, for instance whether certain mutations influence the transport of antibiotics in a bacterial porin and to analyze the gating dynamics of the translocase channel Tim23.

\section*{Acknowledgment}
Financial support of DFG (CRC803, projects Z02 and A01) and of the Volkswagen Foundation (FBMS) is gratefully acknowledged. O.M.S. thanks the GGNB PBCS for a Ph.D. fellowship. We thank Timo Aspelmeier, Annika Bartsch, Niels Denkert, Manuel Diehn, Thomas Hotz, Housen Li, Michael Meinecke, Ingo P. Mey, Erwin Neher, Dominic Schuhmacher, Ivo Siekmann and Frank Werner for fruitful discussions. We also thank two anonymous referees for their helpful comments and suggestions which helped us to improve this paper.

\newpage
\bibliographystyle{plainnat}
\bibliography{MultiscaleBib}

\end{document}